\documentclass{article}
\usepackage{hyperref}
\pdfoutput=1
\usepackage[in]{fullpage}
\usepackage[ascii]{inputenc}
\usepackage[english]{babel}
\usepackage{color}
\usepackage{array}
\usepackage{hhline}
\usepackage{hyperref}
\usepackage[square,sort,comma,numbers]{natbib}
\usepackage[pdftex]{graphicx}
\usepackage{alltt}
\usepackage{listings}
\usepackage{placeins}

\definecolor{lgray}{gray}{0.50}
\hypersetup{colorlinks=true,linkcolor=blue,citecolor=blue,filecolor=blue,urlcolor=blue,
pdftitle={Implementing a Case Management Modeling and Notation (CMMN) System using a Content Management Interoperability Services (CMIS) compliant repository},
pdfauthor={Mike A. Marin, and Jay A. Brown},
pdfsubject={Computer Science, Information systems},
pdfkeywords={Case Handling, Case Management, Case Management System, Case Management Modeling and Notation, CMMN, CMMN Implementation, Content Management, Content Management System, Content Management Interoperability Services, CMIS}}

\newcounter{Figure}

\title{Implementing a Case Management Modeling and Notation (CMMN) System using a Content Management Interoperability Services (CMIS) compliant repository}
\author{Mike A. Marin$^{1,2}$, and Jay A. Brown$^{1}$\\
$^{1}$ IBM Analytics Group\\
3565 Harbor Blvd., Costa Mesa, CA 92626 U.S.A.\\
\texttt{\{mikemarin,jay.brown\}@us.ibm.com}\\
$^{2}$ University of South Africa\\
Pretoria, South Africa
}
\date{\today}
\begin{document}
\maketitle
\begin{abstract}
This paper describes how a Case Management Modeling and Notation (CMMN)
implementation can use Content Management Interoperability Services (CMIS) to implement the CMMN information model. 
The interaction between CMMN and CMIS is described in detail, and two implementation alternatives are presented. 
An \textit{integration} alternative where any external CMIS repository is used. 
This alternative is useful to process technology vendors looking to integrate with CMIS compliant repositories.
An \textit{embedded} alternative where a CMIS repository is embedded within the CMMN engine. 
This alternative is useful to content management vendors implementing CMMN.
In both alternatives a CMIS folder is used as the case file containing the case instance data.
The CMIS repository can also be used to store the CMMN models to take advantage of CMIS versioning and meta-data.
Extensive Java pseudocode is provided as an example of how a CMMN implementation can use a CMIS repository to implement the CMMN information model.
No extensions to CMIS are needed, and only minor extensions to CMMN are proposed.

\end{abstract}


\section{Introduction}

In May 2014, the Object Management Group (OMG) formally released version 1.0 of the Case Management Modeling and Notation (CMMN) \cite{Omg2014cmmn} standard specification. 
The specification is intended to support case management applications  \cite{Marin2013cmmn}.
CMMN is based on two models, a behavioral model and an informational model. The
CMMN specification indicates that the information model can be
implemented using the Content Management Interoperability Services
(CMIS) \cite{OASIS2012cmis} specification, however no
 details are given. This paper addresses that gap by describing how an CMMN
implementation can use CMIS effectively. 
This paper is intended for implementors of CMMN, and should be read in conjunction with the CMMN specification \cite{Omg2014cmmn} and the CMIS specification \cite{OASIS2012cmis}. 
Familiarity with the CMMN and CMIS specifications is assumed.

Case management \cite{Clair2009,Swenson2010,Hill2012} is intended to support the needs of knowledge workers when engaged in knowledge intensive goal oriented processes. 
It is common for knowledge workers to interact via documents (e.g. text documents, word processor documents, spreadsheets, presentations, correspondence, memos, videos, pictures, etc.).
Case management shares most of the knowledge intensive processes characteristics as defined by Di Ciccio \textit{et. al.} which are knowledge driven, collaboration oriented, unpredictable, emergent, goal oriented, event driven, constraint and rule driven, and non repeatable \cite{Ciccio2015}. 
Therefore, it makes sense that a platform to support knowledge workers provide content management and collaboration capabilities. 
Case management is defined by Forrester as:
\begin{quotation}
''A highly structured, but also collaborative, dynamic, and information-intensive process that is driven by outside events and requires incremental and progressive responses from the business domain handling the case. Examples of case folders include a patient record, a lawsuit, an insurance claim, or a contract, and the case folder would include all the documents, data, collaboration artifacts, policies, rules, analytics, and other information needed to process and manage the case.'' \cite{Clair2009}
\end{quotation} 

This paper starts with a short introduction to CMMN in section~\ref{sec:Main-CMMN} and CMIS in section~\ref{sec:Main-CMIS}. 
These introductions describe the main concepts, classes, and objects that will be used in the rest of the paper.
Section~\ref{sec:Main-Alternatives} describes the two implementation alternatives. 
Section~\ref{sec:Main-Interaction} describes how the CMMN information model could be implemented in a CMIS repository. 
Section~\ref{sec:Main-Models} describes the implications for the CMMN models and for process interchange of case models.
An example is given in Section~\ref{sec:Main-Example}. 
The example describes some of the functionality the end users will observe in a CMMN implementation that uses a CMIS repository as described in this paper.
Conclusions are presented in section~\ref{sec:Main-Conclusions}.
Two appendixes are included. 
Appendix~\ref{sec:App-metamodels} shows the CMMN and CMIS meta-models for reference purposes.
Finally, appendix~\ref{sec:App-pseudocode} provides an example Java pseudocode showing a possible implementation of the CMMN information model in CMIS. 

\begin{table}
\centering
\begin{tabular}{|l|l|} \hline
\textbf{CMMN information model class} & \textbf{Corresponding CMIS class}\\\hline
CaseFile & cmis:folder\\\hline
CaseFileItem & cmis:object\\\hline
CaseFileItemDefinition & cmis:object Type\\\hline
Property & cmis:property Type\\\hline
\end{tabular}
\caption{Mapping CMMN information model to CMIS meta-model}
\label{table:Mapping}
\end{table}

\section{Case Management Modeling and Notation (CMMN)}
\label{sec:Main-CMMN}

The CMMN information model starts with a \texttt{CaseFile} that contains \texttt{CaseFileItem}s.
The important classes in the information model (see Figure~\ref{fig:CMMNcasefile}) are,
\begin{description}
\item[\texttt{CaseFile}:] The container for all the information in a case instance. 
The information in a \texttt{CaseFile} can be structured like discrete properties and variables, or unstructured like documents, pictures, voice recordings, video clips, etc.
There is a single \texttt{CaseFile} in a case instance.
It seems natural to implement the \texttt{CaseFile} as a folder (or directory) in a content management system (or file system). 
\item[\texttt{CaseFileItem}:] A piece of information in a case instance. 
All the \texttt{CaseFileItem}s of a case instance are stored in the case's \texttt{CaseFile}.
A case instance may have a large number of \texttt{CaseFileItem}s. 
When using content management system (or file system), it seems natural to implement each \texttt{CaseFileItem} as a document or a folder. 
In most content management systems, both folders and documents have properties that can be used to store structured information. For example, a folder could be used to represent a customer. That folder may have properties like the name, customer number,  phone number, physical and email address of the customer, etc. The folder may be used to store all the emails and documents related with that customer. That folder and all its information maybe part of a case instance, and so stored in the \texttt{CaseFile}. 
\item[\texttt{CaseFileItemDefinition}:] Corresponds to the type of a \texttt{CaseFileItem}.
\item[\texttt{Property}:] Corresponds to a property or field of a \texttt{CaseFileItem}.
A \texttt{CaseFileItem} may have many properties.
In a content management system, properties are often referred as the meta-data of the documents or folders in the system.
\end{description}
The CMMN information model is shown in Figure~\ref{fig:CMMNhighlevel} and Figure~\ref{fig:CMMNcasefile}. 
Figure~\ref{fig:CMMNhighlevel} shows the high level case model, and Figure~\ref{fig:CMMNcasefile} shows the details of the case file model. 
Note that a \texttt{CaseFileItem} has two self-referencing relationships:
\begin{itemize}
\item A composition relationship between \texttt{parent} and \texttt{children} that can be used to represent a folder structure, where the folder (\texttt{CaseFileItem}) contains either documents (other \texttt{CaseFileItem}s) or other folders (also \texttt{CaseFileItem}s).
\item A reflexive association between \texttt{sourceRef} and \texttt{targetRef} that can be used to represent relationships between documents or folders.
\end{itemize}

\begin{figure}
\centering
\includegraphics[keepaspectratio,width=4in]{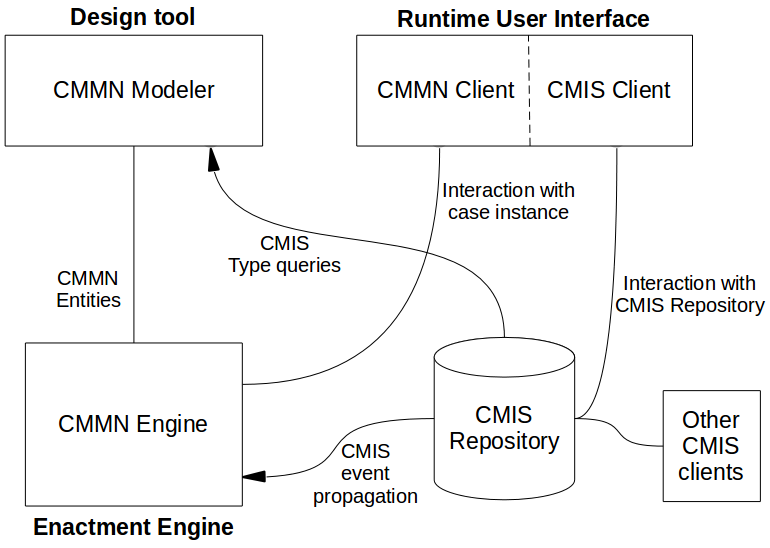}
\caption{Integration alternative}
\label{fig:IntegrationOp}
\end{figure}

\subsection{Implementation using CMIS}

In CMIS, we will use a folder to represent the case instance's \texttt{CaseFile}, and it will contain all the case \texttt{CaseFileItem}s for that case instance.
The \texttt{CaseFileItem}s will be documents or other folders. 
In CMIS as in CMMN, \texttt{CaseFileItem}s (documents or folders) are typed.
A \texttt{CaseFileItem} is an instance of a \texttt{CaseFileItemDefinition}. 
This model can be easily mapped into CMIS as described in Table~\ref{table:Mapping}.

Although there are multiple alternatives to implement the CMMN information model using CMIS, this paper explores just two alternatives:
\begin{itemize}
\item The \textit{integration} alternative where an external CMIS repository is used. 
This alternative will be attractive to process technology vendors that want their technology to integrate with one or more existing CMIS compliant repositories.
\item The \textit{Embedded} alternative where a CMIS repository is embedded within the CMMN engine. 
This alternative will be attractive to content management vendors implementing CMMN over their CMIS compliant repository.
\end{itemize}

\begin{figure}
\centering
\includegraphics[keepaspectratio,width=4in]{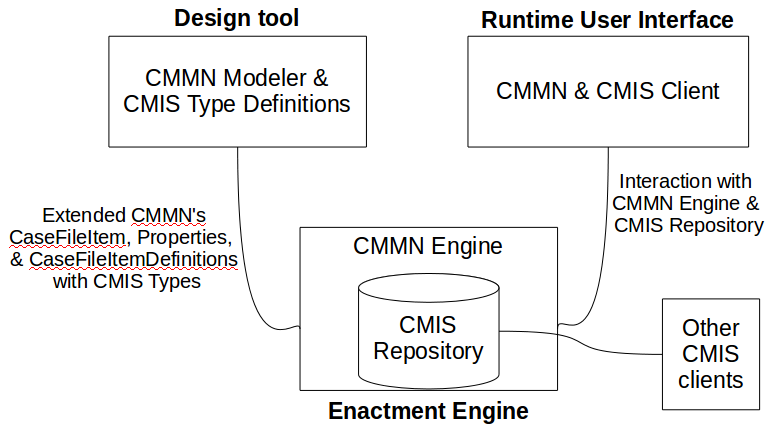}
\caption{Embedded alternative}
\label{fig:EmbeddedOp}
\end{figure}

In both cases, the CMIS repository is used to store all or part of the CMMN information model. 
In both cases, the design tool could create CMIS declarations (mutable types), and the runtime user interface may provide access to both the CMMN engine and the CMIS repository.
Figures \ref{fig:IntegrationOp}, and \ref{fig:EmbeddedOp} show a high level view of the two options.

\begin{figure}
\centering
\includegraphics[keepaspectratio,width=6in]{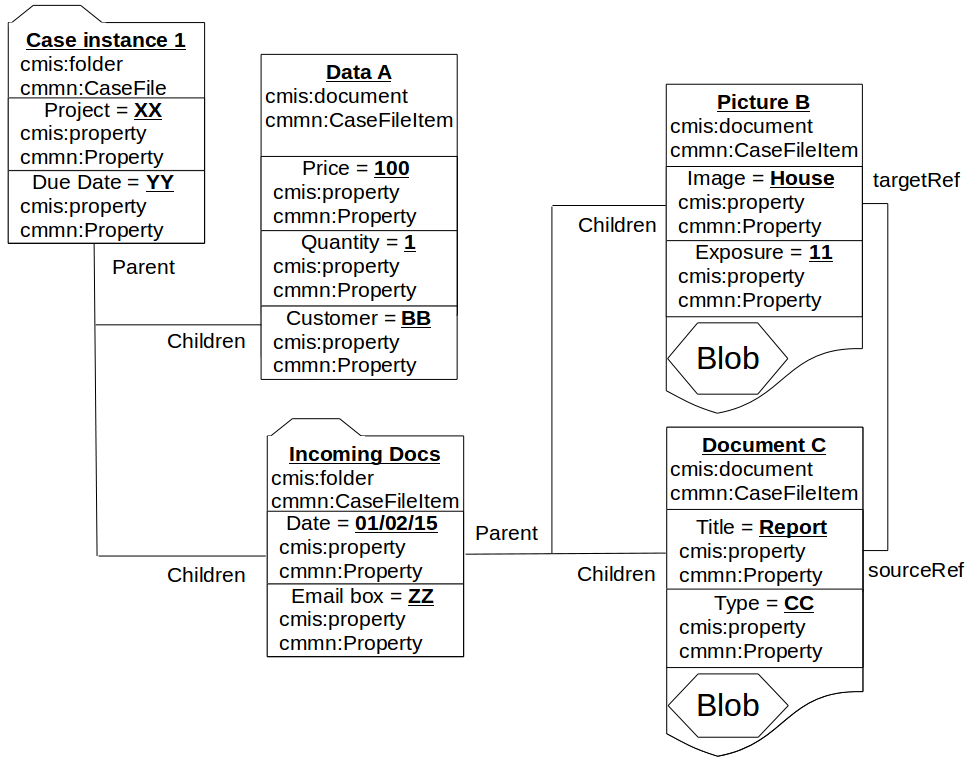}
\caption{Example of a case instance in a CMIS repository}
\label{fig:Example}
\end{figure}

An example of a simple case instance is shown in Figure~\ref{fig:Example}. 
There are five entities in the figure, two \texttt{cmis:folder}s and three \texttt{cmis:document}s. 
Each entity starts with three text lines. 
The first line indicates the name of the entity.
The second line indicates the CMIS object that implements the entity, and the third line indicates the CMMN object that is being implemented.
For illustration purposes, each entity has two or three properties.
The example shows a case file instance (\texttt{CaseFile}) with four \texttt{CaseFileItem}s. 
Data A is a \texttt{cmis:document} that is being used for structured data so it has no document content (blob). 
From a CMIS perspective, Data A is a \texttt{cmis:document} that is missing the \texttt{ContentStream} (See the CMIS meta-model in Figure~\ref{fig:CMISmetamodel}).
This type of documents are normally called content less documents, and they could be implemented as \texttt{cmis:item} instead of \texttt{cmis:document}.
Incoming documents is a folder used to store picture B and document C.
Both picture B and document C are real documents with blobs. 
Picture B is an image of a house, and document C is a report.
There is a relationship between document C and picture B, as probably picture B is mentioned in the report.

For simplicity purposes this paper assumes the full CMMN information model is stored in a CMIS repository. 
Some implementations may decide to implement part of the information model in another database.
The Java pseudocode presented in this paper is intended as an example on how a CMMN implementation may access CMIS, and it is not intended to be used as is.
It is assumed that \texttt{CaseFile} can contain properties as any other \texttt{cmis:folder}, which is not allowed in CMMN.
Hints on how to implement security, versions, and the mapping of CMIS events to CMMN events are given but not fully described in the Java pseudocode.

\section{Content Management Interoperability Services (CMIS)}
\label{sec:Main-CMIS}

The CMIS specification is an open standard for dealing with Enterprise Content Management (ECM) repositories. 
It defines a common domain model and a set of three protocol bindings, each exposing the same domain model but serving a different type of client (SOAP, AtomPub, and JSON). 
OASIS approved CMIS as an official Specification in May 2010  and  
CMIS 1.1 was approved in December 2012 \cite{OASIS2012cmis} and it remains the current version of the specification. The specification states:

\begin{quote}
''The CMIS interface is designed to be layered on top of existing Content Management systems and their existing programmatic interfaces. It is not intended to prescribe how specific features should be implemented within those CM systems, nor to exhaustively expose all of the CM system's capabilities through the CMIS interfaces. Rather, it is intended to define a generic/universal set of capabilities provided by a CM system and a set of services for working with those capabilities.'' \cite{OASIS2012cmis}
\end{quote}

The CMIS domain model consists of a set of nine services (below) and a data model (also discussed below).  It does not attempt to address all aspects of a typical ECM  repository (e.g. Administrative functions, workflow, etc.) rather just the functions typically used at runtime by ECM client applications.   The model's  nine services are detailed below.

\subsection{CMIS Services and Data Model}

CMIS defines a set of 9 services which include:

\begin{description}
\item[Repository Services:] Used to discover info and capabilities of the connected repository.
\item[Navigation Services:] Used to traverse the repository's folder hierarchy.
\item[Object Services:] Used to perform Create Read Update and Delete functions on objects.
\item[Multi-Filing Services:] Allows for multi-filing documents (not folders) within folders. 
\item[Discovery Services:] Exposes Query (based on SQL 92) and getChanges which returns accumulated changes to the repository for indexers. 
\item[Versioning Services:] Used to checkout documents and work with document versions.
\item[Relationship Services:] Used to discover and manage an object's relationships. 
\item[Policy Services:] Used to apply, remove, and query for policies.
\item[ACL Services:] Used to discover and manipulate ACLs (and ACEs) on an object.
\end{description}

The CMIS data model consists of a Repository object that reports all of the capabilities of the ECM repository,  and a hierarchy of objects that are stored in the repository (see Figure~\ref{fig:CMISmetamodel}).  
Each of these objects has a corresponding type or property definition (Shown in Figure~\ref{fig:CMIStypes}).
The important objects of the data model are,
\begin{description}
\item[\texttt{cmis:object}:] Base for \texttt{cmis:document}, \texttt{cmis:folder}, \texttt{cmis:relationship}, \texttt{cmis:policy}, and \texttt{cmis:item}. 
\item[\texttt{cmis:document}:] Describes documents in a content management system. Documents can be of any type, including pictures, video, voice recordings, work processor documents, spreadsheets, etc.
\item[\texttt{cmis:folder}:] Describes folders in a content management system. CMIS supports a hierarchical folder structure.
\item[\texttt{cmis:relationship}:] Can be used to establish a relationship between any two objects.
\item[\texttt{cmis:policy}:] Is an object that can be applied to other \texttt{cmis:object}s. 
The CMIS specification does not describes any specific behavior for these objects.
\item[\texttt{cmis:item}:]  Used to model other object types that does not fit document, folder, relationship or policy types.
An example could be content less documents.
\item[\texttt{cmis:secondary}:] Define a set of properties that can be dynamically applied to an object.
They can be used as markers to create dynamic collections of objects.
\end{description}

CMIS implementations exist for the most popular platforms and languages in Apache Chemistry \cite{Chemistry2014}.  These include libraries for Java, .Net, Python, JavaScript, PHP, Android and iOS (Objective-C).  Finally a complete client development guide \cite{Muller2013}  and a server development guide \cite{Brown2014} are available in addition to all the materials available on the Apache Chemistry site \cite{Chemistry2014}. 

\begin{figure}
\centering
\includegraphics[keepaspectratio,width=5in]{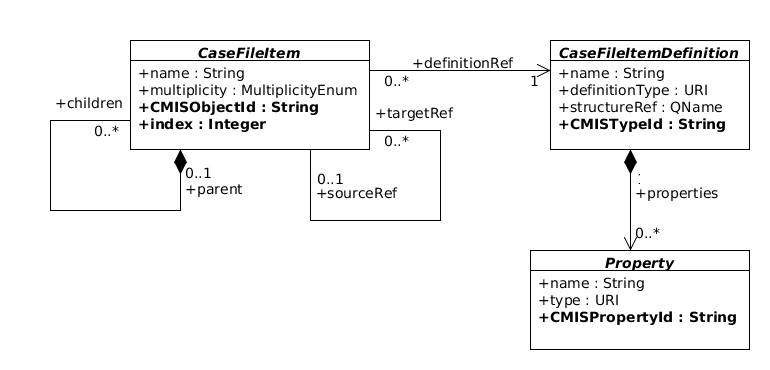}
\caption{Extending CMMN meta-model for CMIS Integration (integration alternative)}
\label{fig:CombinedIntegrated}
\end{figure}

\begin{figure}
\centering
\includegraphics[keepaspectratio,width=6.5in]{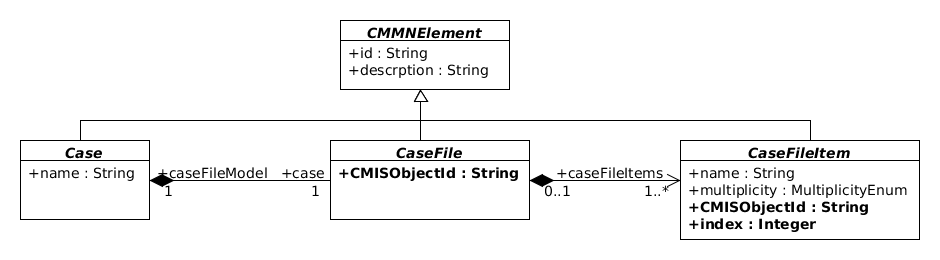}
\caption{Integrated CaseFile}
\label{fig:CaseFileIntegrated}
\end{figure}

\section{Alternatives}
\label{sec:Main-Alternatives}

In this section the two alternatives are described, \textit{integration} and \textit{embedded} alternatives. 
The \textit{integration} alternative will be appealing to process technology vendors, because it allows their CMMN implementations to use external CMIS repositories from other vendors. 
This may allow those process technology vendors to support one or more CMIS compliant repositories. 
The \textit{embedded} alternative will be appealing to content management vendors, because it allows them to implement CMMN within their CMIS compliant repositories.
Content management vendors entering the process space may prefer the \textit{embedded} alternative.

 \begin{figure}
\centering
\includegraphics[keepaspectratio,width=6.5in]{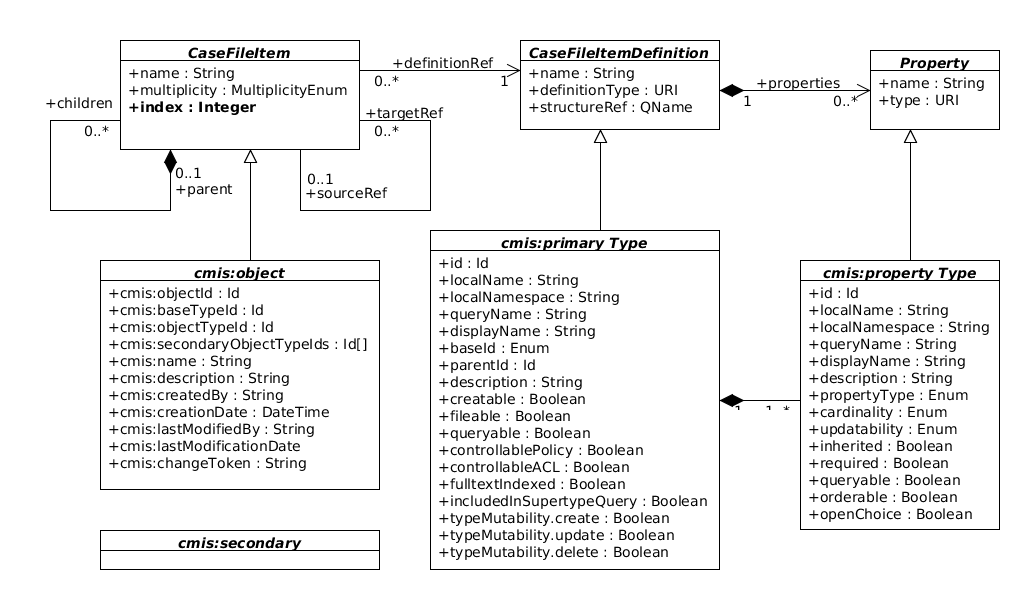}
\caption{Extending CMMN meta-model for embedding CMIS (embedded  alternative)}
\label{fig:CombinedEmbeded}
\end{figure}

\subsection{Integration alternative}

In this alternative the CMIS repository is external to the CMMN engine, and it could be implemented using any CMIS 1.1 compliant repository.
This may allow a CMMN implementation to be compatible with one or more CMIS 1.1 repositories.
Figure \ref{fig:IntegrationOp} shows a high level view of the integration alternative showing the touch points required to implement it. 
The external CMIS repository is used to store all or part of the CMMN information model. 

Figure~\ref{fig:CombinedIntegrated} shows the CMMN
meta-model extended to integrate with a CMIS repository, and  
Figure~\ref{fig:CaseFileIntegrated} shows a case file
implemented by a \texttt{cmis:folder}. To support the \textit{integration} alternative, the CMMN meta-model needs to be enhanced by adding references to the CMIS objects, as follows
\begin{itemize}
\item The \texttt{CaseFile} now has a \texttt{CMISObjectId} attribute to reference the  \texttt{cmis:folder}'s \texttt{Id}  that implements the case instance in the CMIS repository.
\item The \texttt{CaseFileItem} now has a \texttt{CMISObjectId} attribute to reference the corresponding \texttt{cmis:object}'s \texttt{Id} in the CMIS repository. In addition, it also has a \texttt{index} to implement the multiplicity concept in CMMN.
\item The \texttt{CaseFileItemDefinition} now has a \texttt{CMISTypeId} attribute to indicate the \texttt{Id} of a  \texttt{cmis:object Type}.
\item The \texttt{Property} now has a \texttt{CMISPropertyId} attribute to indicate the \texttt{Id} of a \texttt{cmis:property Type}
\end{itemize}

The CMMN design tool and the CMMN runtime client user interface may or may not be aware of the CMIS repository.
An implementation in which the CMMN design tool is aware of the CMIS repository may allow users to create CMIS declarations (mutable types) corresponding to different document or folder types (\texttt{CaseFileItem} types' \texttt{CaseFileItemDefinition}s). 
Because both \texttt{cmis:folder}s and \texttt{cmis:documents} can contain properties, the complete CMMN information model can be implemented in CMIS.
For example an implementation may use properties in the \texttt{cmis:folder} to store case \texttt{CaseFile} properties.

The runtime user interface could include a CMIS client, giving the end users the ability to inspect and modify all the cases (\texttt{CaseFile}'s \texttt{cmis:folder}s) based on his or her level of security access. When an end user modifies a case (\texttt{cmis:folder}) by adding documents, folders, or modifying the case folder documents or folders or their properties. The corresponding events are raised by CMIS and the CMMN implementation should react by evaluating and triggering the correct sentries' \texttt{onPart}s. 

\begin{figure}
\centering
\includegraphics[keepaspectratio,width=6.5in]{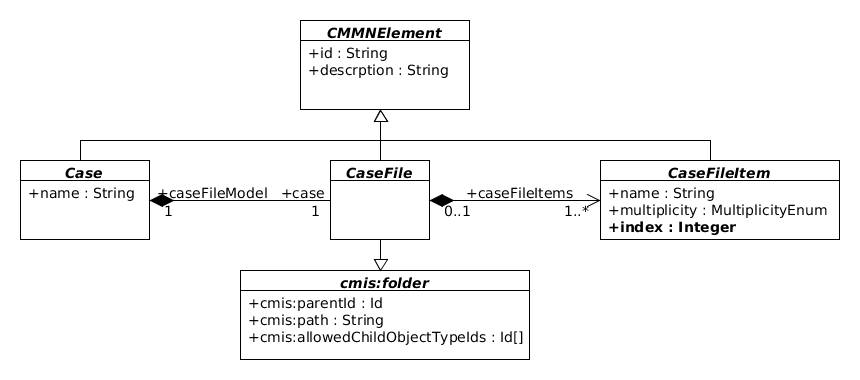}
\caption{Embedded CaseFile}
\label{fig:CaseFileEmbeded}
\end{figure}

\subsection{Embedded  alternative}

The \textit{embedded} alternative is the implementation of both CMIS and CMMN in the same engine. 
In this alternative, the CMIS repository is embedded within the CMMN engine. 
Figure \ref{fig:EmbeddedOp} shows a CMIS repository embedded in a CMMN implementation. 

Figure \ref{fig:CombinedEmbeded} describes the merged
meta-model between CMMN and CMIS. 
Figure \ref{fig:CaseFileEmbeded} shows the case file implemented by a \texttt{cmis:folder}.
To support the \textit{embedded} alternative, the CMMN meta-model needs to be enhanced as follows,
\begin{itemize}
\item The \texttt{CaseFile} becomes a generalization of  \texttt{cmis:folder}.
\item The \texttt{CaseFileItem} becomes a generalization of a \texttt{cmis:object}, and now has a \texttt{index} to implement the multiplicity concept in CMMN.
\item The \texttt{CaseFileItemDefinition} becomes a generalization of \texttt{cmis:Object Type}.
\item The \texttt{Property} becomes a generalization of \texttt{cmis:property Type}
\end{itemize}

Everything that can be implemented with the \textit{integration} alternative is also possible in \textit{embedded} alternative. 
In addition, the \textit{embedded} alternative provides advantages over the \textit{integration} alternative. 
In particular the CMMN and CMIS design and runtime information may be stored in the same database.
For example, event propagation can be done more efficiently, because a push model could be implemented versus the pull model described in this paper and in the Java pseudocode section~\ref{sec:event}.

\section{A CMIS repository as the CMMN information model}
\label{sec:Main-Interaction}

Independent of the alternative implemented (\textit{integration} or \textit{embedded}) the interaction between CMMN and
CMIS will follow similar patterns and the CMMN implementation will use similar CMIS APIs. 
The simple way to use the CMIS API is to use one of the Apache Chemistry libraries \cite{Chemistry2014}.
Appendix~\ref{sec:App-pseudocode} \nameref{sec:App-pseudocode} shows Java pseudocode using the OpenCMIS Java API from Apache Chemistry \cite{OpenCMIS2015} to describe at a high level how a CMMN implementation can invoke CMIS functionality.

It is important to notice that the CMMN meta-model in Figures~\ref{fig:CMMNhighlevel} and~\ref{fig:CMMNcasefile} describe a modeling time meta-model that can be used for process interchange. 
While the CMIS meta-model in Figure~\ref{fig:CMISmetamodel} is a runtime model representing objects in a content management repository. 
The CMIS types meta-model in Figure~\ref{fig:CMIStypes} can be considered both a modeling and runtime meta-model.
In here we combine all of those two meta-models and use them for both modeling and runtime execution.

\begin{table}
\centering
\begin{tabular}{|p{2.7cm}|p{3cm}|p{9cm}|} \hline
\textbf{CMIS \newline Object-type} &
\textbf{CMMN \newline Definition Type} &
\textbf{CMMN CaseFileItemDefinition Type's URI}\\\hline
\ cmis:folder  &
CMIS Folder  &
\verb|http://www.omg.org/spec/CMMN/DefinitionType/| \newline \verb|CMISFolder|\\\hline
\ cmis:document  &
CMIS Document &
\verb|http://www.omg.org/spec/CMMN/DefinitionType/| \newline \verb|CMISDocument|\\\hline
\ cmis:relationship  &
CMIS \newline Relationship &
\verb|http://www.omg.org/spec/CMMN/DefinitionType/| \newline \verb|CMISRelationship|\\\hline
\ {}-{}- &
XML-Schema \newline Element &
\verb|http://www.omg.org/spec/CMMN/DefinitionType/| \newline \verb|XSDElement|\\\hline
\ {}-{}- &
XML Schema \newline Complex Type &
\verb|http://www.omg.org/spec/CMMN/DefinitionType/| \newline \verb|XSDComplexType|\\\hline
\ {}-{}- &
XML Schema \newline Simple Type &
\verb|http://www.omg.org/spec/CMMN/DefinitionType/| \newline \verb|XSDSimpleType|\\\hline
\ {}-{}- &
Unknown &
\verb|http://www.omg.org/spec/CMMN/DefinitionType/| \newline \verb|Unknown|\\\hline
\ {}-{}- &
Unspecified &
\verb|http://www.omg.org/spec/CMMN/DefinitionType/| \newline \verb|Unspecified|\\\hline 
\multicolumn{3}{ |c| }{\textbf{Extend CMMN as follows}}\\\hline
\ cmis:policy &
\ {}-{}- &
\verb|http://www.omg.org/spec/CMMN/DefinitionType/| \newline \verb|CMISPolicy|\\\hline
\ cmis:item &
\ {}-{}- &
\verb|http://www.omg.org/spec/CMMN/DefinitionType/| \newline \verb|CMISItem|\\\hline
\ cmis:secondary  &
\ {}-{}- &
\verb|http://www.omg.org/spec/CMMN/DefinitionType/| \newline \verb|CMISSecondary|\\\hline
\multicolumn{3}{ |c| }{\textbf{other CMIS Object-types}}\\\hline
\ cmis:object  &
\textit{any of:} \newline
CMIS Folder  \newline \newline
CMIS Document\newline \newline
CMIS \newline Relationship \newline
\ {}-{}- \newline \newline
\ {}-{}- \newline \newline
\ {}-{}- \newline  &
\ \newline
\verb|http://www.omg.org/spec/CMMN/DefinitionType/| \newline \verb|CMISFolder| \newline
\verb|http://www.omg.org/spec/CMMN/DefinitionType/| \newline \verb|CMISDocument| \newline
\verb|http://www.omg.org/spec/CMMN/DefinitionType/| \newline \verb|CMISRelationship| \newline
\verb|http://www.omg.org/spec/CMMN/DefinitionType/| \newline \verb|CMISPolicy| \newline
\verb|http://www.omg.org/spec/CMMN/DefinitionType/| \newline \verb|CMISItem| \newline
\verb|http://www.omg.org/spec/CMMN/DefinitionType/| \newline \verb|CMISSecondary|\\\hline
\end{tabular}
\caption{Object Types}
\label{table:objTypes}
\end{table}

\subsection{CMMN Information Model}

The information model in CMMN is based on a \texttt{CaseFile} (see
Figure~\ref{fig:CMMNhighlevel} and
Figure~\ref{fig:CMMNcasefile}). When implementing
using CMIS, there are at least two options that can be used to implement the \texttt{Casefile}, 
\begin{description}
\item[Use a \texttt{cmis:folder}] to represent the \texttt{CaseFile}.
An \textit{integration} alternative, as shown on Figure~\ref{fig:CaseFileIntegrated}, may include a \texttt{CMISObjectId} in the \texttt{CaseFile} as a reference to the \texttt{cmis:folder} representing the case instance in CMIS.
An \textit{embedded} alternative, as shown in Figure~\ref{fig:CaseFileEmbeded}, may implement the \texttt{CaseFile} as a generalization of a \texttt{cmis:folder}.
Each case instance will have a \texttt{cmis:folder} containing all the case file items for that case. 
A \texttt{cmis:folder} can contain properties, and so, case properties can also be implemented in the \texttt{cmis:folder} implementing the \texttt{CaseFile}.
The \texttt{cmis:folder} will probably
outlive the case instance lifecycle, which is a good side effect,
because all the case file items for a case instance will remain in the
\texttt{cmis:folder} after the case is completed. 
\item[Use a database] to implement the \texttt{Casefile}, with references to its content in the CMIS repository. 
Under this
option, there is no CMIS representation of the case file, and so, 
the implementation will need to keep track of the CMIS objects stored in the
case (most likely documents and folders) by storing their \texttt{cmis:objectId} in the database.
\end{description}

This paper describes the first option of using a \texttt{cmis:folder} to represent the \texttt{CaseFile}.
The CMMN information model matches well to the CMIS model. In both meta-models, CMMN and CMIS,
there is a class that represents an object, a class that represents the
type of that object, and a property class. 
 Therefore, we can map 
between the two specifications as shown in Table~\ref{table:Mapping}. 
All the content in a CMIS repository can be represented by \texttt{cmis:object}s and their descendants. Similarly in CMMN the information model is represented by \texttt{CaseFileItem}s. 
Therefore, we can map \texttt{CaseFileItem}s to \texttt{cmis:object}s. 
That allows the \texttt{CaseFileItem}s to describe all the CMIS objects, including documents and folders.
Note that in CMMN, \texttt{CasefileItem}s representing folders use the \texttt{children} relationship (see Figure~\ref{fig:CMMNcasefile}) to point to the \texttt{CaseFileItem}s stored in the folder. 
As we described before, a \texttt{CaseFile} can be mapped to a \texttt{cmis:folder}.
A \texttt{CaseFileItemDefinition} naturally maps into \texttt{cmis:object Type},
because \texttt{CaseFileItemDefinition} describes the type of a \texttt{CaseFileItem}, and so, it is similar to a \texttt{cmis:object Type} which
defines the type of a \texttt{cmis:object}.
\texttt{Property} and \texttt{cmis:property Type} represent the same concept.
Therefore, the mapping in Table~\ref{table:Mapping} allow us to describe the full CMMN information model using CMIS. 

The only high level CMIS objects not included in Table~\ref{table:Mapping} are \texttt{cmis:policy}, \texttt{cmis:item}, and \texttt{cmis:secondary}. 
They are optional in CMIS and probably not required for most CMMN implementations. 
However, below we describe how they could be implemented if needed, by indirectly mapping them to \texttt{CaseFileItem}s (see Table~\ref{table:objTypes}).  

We use \texttt{cmis:folders} to implement the \texttt{CaseFileItem} self-referencing composition relationship between \texttt{parent} and \texttt{children} (see Figure~\ref{fig:CMMNcasefile}). 
We use \texttt{cmis:relationship} to implement the \texttt{CaseFileItem} self-referencing reflexive association between \texttt{sourceRef} and \texttt{targetRef} (see Figure~\ref{fig:CMMNcasefile}).

\subsubsection{Objects and data types}

In CMMN a \texttt{CaseFileItem} can represent many objects, and the
\texttt{CaseFileItemDefinition} defines the type by using a URI. 
In CMIS each
object has its own class that is a specialization of \texttt{cmis:object}.
Therefore to represent a CMIS object in CMMN, we need to set the
correct URI value in the CMMN's \texttt{CaseFileItemDefinition}'s
\texttt{definitionType}, while assigning the correct CMIS object via \texttt{cmis:object}
specialization to the \texttt{caseFileItem}.

\paragraph{Objects} \ 

Table~\ref{table:objTypes} compares the object types in CMIS with the CMMN information model. 
The object types in CMMN are defined by the \texttt{DefinitionTypeEnum}.
The CMMN's \texttt{CaseFileItemDefinition} describes the type of the \texttt{CaseFileItem} using an URI that includes some CMIS types. 
Table~\ref{table:objTypes} does an explicit mapping between CMIS object types and CMMN's \texttt{CaseFileItemDefinition} types. 
The CMIS types described by the CMMN's URIs should be enough for most implementations, but if needed three more URIs can be added for \texttt{cmis:policy}, \texttt{cmis:item}, and \texttt{cmis:secondary}, as described in Table~\ref{table:objTypes}. 
Note that CMIS policy, item, and secondary are optional in CMIS, and some implementions may not implement them.

\paragraph{Data types} \ 

CMMN and CMIS property types are based on the XML Schema types \cite{W3C2004Schema}.
CMMN uses most of the XML Schema types, while CMIS uses a limited set of types. 
This makes it easy to map CMIS types to CMMN. 
Table~\ref{table:fieldTypes} maps the CMIS types onto CMMN types. 
To fully support CMIS, the CMMN property \texttt{type} URI needs to be extended
with \texttt{xsd:decimal}, \texttt{Id}, and \texttt{HTML}.

\begin{table}
\centering
\begin{tabular}{|l|l|l|} \hline
\textbf{Type} &
\textbf{CMIS Type} &
\textbf{CMMN Property Type's URI}\\\hline
string & xsd:string & \verb|http://www.omg.org/spec/CMMN/PropertyType/string|\\\hline
boolean & xsd:boolean & \verb|http://www.omg.org/spec/CMMN/PropertyType/boolean|\\\hline
integer & xsd:integer & \verb|http://www.omg.org/spec/CMMN/PropertyType/integer|\\\hline
float & {}-{}- & \verb|http://www.omg.org/spec/CMMN/PropertyType/float|\\\hline
double & {}-{}- & \verb|http://www.omg.org/spec/CMMN/PropertyType/double|\\\hline
duration & {}-{}- & \verb|http://www.omg.org/spec/CMMN/PropertyType/duration|\\\hline
dateTime & xsd:dateTime & \verb|http://www.omg.org/spec/CMMN/PropertyType/dateTime|\\\hline
time & {}-{}- & \verb|http://www.omg.org/spec/CMMN/PropertyType/time|\\\hline
date & {}-{}- & \verb|http://www.omg.org/spec/CMMN/PropertyType/date|\\\hline
gYearMonth & {}-{}- & \verb|http://www.omg.org/spec/CMMN/PropertyType/gYearMonth|\\\hline
gYear & {}-{}- & \verb|http://www.omg.org/spec/CMMN/PropertyType/gYear|\\\hline
gMonthDay & {}-{}- & \verb|http://www.omg.org/spec/CMMN/PropertyType/gMonthDay|\\\hline
gDay & {}-{}- & \verb|http://www.omg.org/spec/CMMN/PropertyType/gDay|\\\hline
gMonth & {}-{}- & \verb|http://www.omg.org/spec/CMMN/PropertyType/gMonth|\\\hline
hexBinary & {}-{}- & \verb|http://www.omg.org/spec/CMMN/PropertyType/hexBinary|\\\hline
base64Binary & {}-{}- & \verb|http://www.omg.org/spec/CMMN/PropertyType/base64Binary|\\\hline
anyURI & xsd:anyURI & \verb|http://www.omg.org/spec/CMMN/PropertyType/anyURI|\\\hline
QName & ~  & \verb|http://www.omg.org/spec/CMMN/PropertyType/QName|\\\hline
\multicolumn{3}{ |c| }{\textbf{Extend CMMN as follows}}\\\hline
{}-{}- & xsd:decimal & \verb|http://www.omg.org/spec/CMMN/PropertyType/decimal|\\\hline
{}-{}- & Id & \verb|http://www.omg.org/spec/CMMN/PropertyType/Id|\\\hline
{}-{}- & HTML & \verb|http://www.omg.org/spec/CMMN/PropertyType/HTML|\\\hline
\end{tabular}
\caption{Property types}
\label{table:fieldTypes}
\end{table}

\subsubsection{Navigating the information model} 

CMMN describes a standard set of seven \texttt{CaseFileItem} operations (see the CMMN specification \cite{Omg2014cmmn} section 7.3.1 CaseFileItem operations) for the behavioral model to navigate the information model. 
The Java pseudocode in section~\ref{sec:ops} \nameref{sec:ops} in 
appendix~\ref{sec:App-pseudocode} \nameref{sec:App-pseudocode} shows a potential implementation of these operations.
All the operations described here work over \texttt{CaseFileItem}s in a case instance.
As described before, all the \texttt{CaseFileItem}s are contained within the \texttt{CaseFile} of the case instance (see Figure~\ref{fig:CMMNcasefile}).
All the operations return ether a \texttt{CaseFileItem} (see Table~\ref{table:objTypes}) instance; or an \texttt{Element} which corresponds to a property of a CMIS object (see Table~\ref{table:fieldTypes}).
Note that to implement CMMN's \texttt{multiplicity} for \texttt{CaseFileItem}s an \texttt{index} has been added to \texttt{CaseFileItem} for both integration or embedded alternatives 
(see Figure~\ref{fig:CaseFileIntegrated}, Figure~\ref{fig:CaseFileEmbeded}, Figure~\ref{fig:CombinedEmbeded}, and Figure~\ref{fig:CombinedIntegrated}). 
The \texttt{index} must be maintained by the implementation and should be incremented when multiple \texttt{cmis:object}s within the same case instance (\texttt{CaseFile}) have the same \texttt{cmis:name}. 
For most implementations that may imply that a CMIS property \texttt{index} must be added to all the \texttt{cmis:object}s that can be stored in a case folder (\texttt{CaseFile}).

The operations defined in the CMMN specification are intended to be used in CMMN expressions. 
Therefore these operations are intended to be implemented as part of the CMMN expression support (see the CMMN specification \cite{Omg2014cmmn} section 5.4.7 Expressions).
The default expression language in CMMN is XPath, however CMMN implementations may support other expression languages. 
Note that an implementation will need to wrap the operations shown in section~\ref{sec:ops} to expose them in the supported expression languages. 
Therefore, the Java pseudocode in appendix~\ref{sec:App-pseudocode} \nameref{sec:App-pseudocode} is intended as an example, and may not implement the CMMN operations exactly as they will be exposed in an expression language.

\subsubsection{Modifying the information model}

The previous section describes how to implement the required CMMN operations to navigate a case instance (\texttt{CaseFile}) information model using the CMIS API.
This section will describe how to use the CMIS API to modify a case instance (\texttt{CaseFile}) information model.
The Java pseudocode in section~\ref{sec:other} \nameref{sec:other} in 
appendix~\ref{sec:App-pseudocode} \nameref{sec:App-pseudocode} shows an implementation of operations to create \texttt{CaseFileItem}s and relationships between them.
Operations to create documents (\texttt{createCaseFileItem DocumentInstance}), folders (\texttt{createCaseFileItemFolderInstance}), and relationships (\texttt{createCaseFileItem Relationship}) are described. 

The \texttt{createCaseFileItemFolderInstance} is used to create \texttt{cmis:folder}s to implement the \texttt{CaseFileItem} self-referencing composition relationship between \texttt{parent} and \texttt{children}. 
The \texttt{createCaseFileItemRelationship} is used to create
\texttt{cmis:relationship}s to implement the \texttt{CaseFileItem} self-referencing reflexive association between \texttt{sourceRef} and \texttt{targetRef}. Those two \texttt{CaseFileItem} self-referencing relationships are shown in Figure~\ref{fig:CMMNcasefile}.

The update and deletion of \texttt{cmis:document}s, \texttt{cmis:folder}s, or \texttt{cmis:relationship}s can be trivially accomplished using Apache Chemistry OpenCMIS \cite{OpenCMIS2015}, with existing method calls and so, are not included in the Java pseudocode.

\begin{figure}
\centering
\includegraphics[keepaspectratio,width=4in]{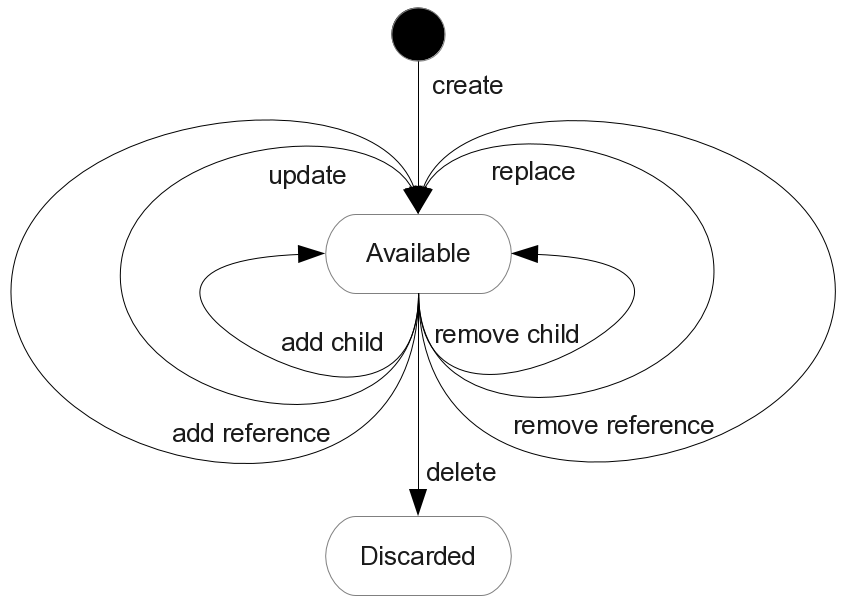}
\caption{CaseFileItem lifecycle}
\label{fig:CMMNFig71}
\end{figure}

\subsection{CaseFileItem Lifecycle event propagation}

For CMMN's sentries to work correctly, events generated
from the CMIS objects must be propagated to the corresponding sentry \texttt{onPart}.
CMMN describes a lifecycle for the \texttt{CaseFileItem} as shown in 
Figure~\ref{fig:CMMNFig71}. 
From a CMIS perspective, we can
separate the lifecycle state transitions between \texttt{cmis:folder}s 
as described in Table~\ref{table:Folder}, and other
CMIS objects (\texttt{cmis:document}, \texttt{cmis:relationship}, \texttt{cmis:policy}, \texttt{cmis:item}, and \texttt{cmis:secondary})
as described in Table~\ref{table:Docs}.

The Java pseudocode in section~\ref{sec:event} \nameref{sec:event} in appendix~\ref{sec:App-pseudocode} \nameref{sec:App-pseudocode} shows how to pull the CMIS repository for events.
Those CMIS events then can be used to evaluate and trigger sentry's \texttt{onPart}.
Calling the \texttt{GetContentChangesForEventPropagation} method in section~\ref{sec:event} \nameref{sec:event}, will place the thread into an infinite loop pulling for CMIS events. 
The implementor will need to complete the \texttt{PushChangeEvents} Java pseudocode method to propagate the events into the CMMN implementation.
OpenCMIS \texttt{Enum ChangeType} has only four values for change events, CREATED, DELETED, SECURITY, and UPDATED. So the developer will have to map these to the the events in Table~\ref{table:Folder} and Table~\ref{table:Docs}.
This exercise is left to the reader.

\subsection{Versioning and Roles}

Although CMIS supports versioning, the CMMN specification states that for purposes of CMMN modeling, only the last version is assumed, but implementations can use versioning if required
(see the CMMN specification \cite{Omg2014cmmn} section 5.3.2.1 Versioning). 
When implementing CMMN using CMIS, it makes sense to take advantage of the CMIS versioning capabilities.

Roles in CMMN are used for human tasks and are not associated with \texttt{CaseFileItems}, however when using CMIS it makes sense to use the CMIS security features to support the CMMN role concept. Each CMIS object (\texttt{cmis:object}) can have a \texttt{ACL} associated with it to implement security.

\begin{table}
\centering
\begin{tabular}{|p{1.2in}|p{1.4in}|p{3.6in}|} \hline
\textbf{CMIS \newline folder event} & \textbf{CMMN \newline CaseFileItem event} & \textbf{Description}\\\hline
file in folder  & addChild & A new object has been added to the folder\\\hline
create relationship & addReference & A new cmis:relationship to the folder has been added\\\hline
create folder  & create & The folder has been created\\\hline
delete folder  & delete & The folder has been deleted\\\hline
unfile document  & removeChild & An object has been removed (un-filed) from the folder\\\hline
delete relationship  & removeReference & A cmis:relationship that pointed to the folder was removed \\\hline
delete + create  & replace & The complete folder was replaced with a new version\\\hline
update folder  & update & The folder properties have been modified\\\hline
\end{tabular}
\caption{CMIS folder events}
\label{table:Folder}
\end{table}

\begin{table}
\centering
\begin{tabular}{|p{1.2in}|p{1.4in}|p{3.6in}|} \hline
\textbf{CMIS document, \newline relationship, \newline policy, item, or secondary event}
& \textbf{CMMN \newline CaseFileItem event} & \textbf{Description}\\\hline
create relationship  & addReference & A new cmis:relationship to the object has been added\\\hline
create  & create & The object has been created\\\hline
delete  & delete & The object has been deleted\\\hline
delete relationship  & removeReference & A cmis:relationship that pointed to the object was removed \\\hline
delete + create  & replace & A new version of the object has replaced the previous version\\\hline
update  & update & The object properties have been modified\\\hline
\end{tabular}
\caption{CMIS document, relationship, policy, item, or secondary events}
\label{table:Docs}
\end{table}

\section{CMMN models}
\label{sec:Main-Models}

This section describes how to store the CMMN models in the CMIS repository. 
It also describe the effects of using CMIS as described in this paper on process interchange.

\subsection{Storing the CMMN Models}

The CMIS repository can be used by the CMMN modeler tool to store the models. 
The modeler tool can take advantage of the versioning offered by most CMIS repositories to maintain the versions of its models.
It can also take advantage of the CMIS folders to create project folders with the ability to create sub-folders to store the multiple assets of a project.
In general, the CMIS repository can be used as the modeler repository for CMMN models and other modeling artifacts.
The CMMN models and other artifacts can be represented as \texttt{cmis:document}s and stored in specialized \texttt{cmis:document Type}s and \texttt{cmis:folder Type}s. 
The CMMN model documents can have specialized meta-data for the CMMN modeler tool to use. 
For example, project name, department, etc.
Standard CMIS meta-data can also be used by the CMMN modeler tool to keep track of its models.
For example, \texttt{cmis:name}, \texttt{cmis:description}, \texttt{cmis:createdBy}, \texttt{cmis:creationDate}, 
\texttt{cmis:lastModifiedBy}, \texttt{cmis:lastModificationDate}, 
\texttt{cmis:versionLabel}, etc.

\subsection{CMMN Extensions}

In order for the CMMN implementation to take full advantage of the capabilities offered by CMIS, few extensions to CMMN are required, as follows. 
\begin{description}
\item[\texttt{Property} types] can be extended as shown in Table~\ref{table:fieldTypes} to support \texttt{xsd:decimal}, \texttt{Id}, and \texttt{HTML} types. Note that if a CMMN application is exclusively using a CMIS repository then it would never encounter one of these types.  So these extensions may be optional.  
\item[\texttt{CaseFileItem} types] may need to be extended as shown in Table~\ref{table:objTypes}. 
This is optional, because not all implementations will need to support all the CMIS objec types. 
Implementations that need to support \texttt{cmis:policy}, \texttt{cmis:item}, or \texttt{cmis:secondary} will need to extend the \texttt{CaseFileItemDefinition definitionType}'s URI as described in Table~\ref{table:objTypes}.
\item[Extended attributes] are needed in both alternatives. 
The \textit{embedded} alternative requires extended attributes to support,
\begin{itemize}
\item \texttt{index} as an attribute of \texttt{CaseFileItem}
\end{itemize} 
The \textit{integration} alternative requires extended attributes to support,
\begin{itemize}
\item \texttt{CMISObjectId} as an attribute of \texttt{CaseFile} and  \texttt{CaseFileItem}
\item \texttt{index} as an attribute of \texttt{CaseFileItem}
\item \texttt{CMISTypeId} as an attribute of \texttt{CaseFileItemDefinition}
\item \texttt{CMISPropertyId} as an attribute of \texttt{Property}
\end{itemize}
\end{description}

In CMMN 1.0, these extensions affect process interchange.
Future versions of the CMMN specification may introduce extensible attributes and rules on how to preserve extended URIs in \texttt{CaseFileItemDefinition definitionType}'s URI and \texttt{Property Type}'s URI.

Currently, tools wishing to preserve CMIS 1.0 process interchange may need to introduce an option when saving CMMN models to indicate if the model must be CMMN 1.0 compatible, and if so, the following transformations will be required, to remove extensions:
\begin{itemize}
\item Remove the extended attributes as follows,
\begin{description}
\item[\texttt{index}] from \texttt{CaseFileItem}
\item[\texttt{CMISObjectId}] from \texttt{CaseFile} and  \texttt{CaseFileItem}
\item[\texttt{CMISTypeId}] from \texttt{CaseFileItemDefinition}
\item[\texttt{CMISPropertyId}] from \texttt{Property}
\end{description}
\item Map extended \texttt{Property Type}s as follows,
\begin{description}
\item[\texttt{xsd:decimal}] (\verb|http://www.omg.org/spec/CMMN/PropertyType/decimal|) to \\ \texttt{double} (\verb|http://www.omg.org/spec/CMMN/PropertyType/double|)
\item[\texttt{xsd:Id}] (\verb|http://www.omg.org/spec/CMMN/PropertyType/Id|) to \\ \texttt{string} (\verb|http://www.omg.org/spec/CMMN/PropertyType/string|)
\item[\texttt{xsd:HTML}] (\verb|http://www.omg.org/spec/CMMN/PropertyType/HTML|) to \\ \texttt{string} (\verb|http://www.omg.org/spec/CMMN/PropertyType/string|)
\end{description}
\item Map extended \texttt{CaseFileItemDefinition definitionType}s as follows,
\begin{description}
\item[\texttt{cmis:policy}] (\verb|http://www.omg.org/spec/CMMN/DefinitionType/CMISPolicy|) to \\ \texttt{Unknown} (\verb|http://www.omg.org/spec/CMMN/DefinitionType/Unknown|)
\item[\texttt{cmis:item}] (\verb|http://www.omg.org/spec/CMMN/DefinitionType/CMISItem|) to \\ \texttt{Unknown} (\verb|http://www.omg.org/spec/CMMN/DefinitionType/Unknown|)
\item[\texttt{cmis:secondary}] (\verb|http://www.omg.org/spec/CMMN/DefinitionType/CMISSecondary|) to \\ \texttt{Unknown} (\verb|http://www.omg.org/spec/CMMN/DefinitionType/Unknown|)
\end{description}
\item Review the generalizations from CMIS classes in the \textit{embedded} alternative, which are,
\begin{description}
\item[\texttt{CaseFile}] generalization of  \texttt{cmis:folder}
\item[\texttt{CaseFileItem}] generalization of  \texttt{cmis:object}
\item[\texttt{CaseFileItemDefinition}] generalization of \texttt{cmis:Object Type}
\item[\texttt{Property}] generalization of \texttt{cmis:property Type}
\end{description}
\end{itemize}

\section{Example}
\label{sec:Main-Example}

This example describes an hypothetical CMMN implementation using a CMIS repository to implement the case file and to store CMMN models, as described in this paper.
In this example, the implementation has two end user front end tools, the modeling tool and the client tool. 
Both front ends may be integrated into a single user interface.
The modeling tool allows users to create CMMN case models, and so, implements the design time aspects of CMMN.
The modeling tool is used by business analysts or case workers to create, update, and manage CMMN models.
Case models are serialized into machine readable files as described in the CMMN specification. 
The files could be XMI or CMMN XML-Schema (XSD) compliant files.
Those files are stored in the CMIS repository as documents.
The client tool allows case workers to interact with a case instance, and so, implements the runtime aspects of a CMMN implementation.
Case workers using the client tool are able to create case instances, interact with case instances by adding content, executing tasks and stages, engaging in planning by adding discretionary items to the case instance plan, collaborating with other case workers to complete case instances, etc.
The case instance information model is implemented in CMIS as \texttt{cmis:folder} representing the case file.
Therefore, each case instance will have its unique CMIS folder.
The user using the client tool can see the state of the case instance in the CMIS folder and associated content.
An example of a case file is shown in Figure~\ref{fig:Example}. 
In that figure, the case instance for project XX has a \texttt{CaseFileItem} Data 1 with some properties, and a sub-folder for incoming documents with two documents, a house picture and a report document.
 
In a system with a clear separation between design and runtime, a business analyst may create a case model and save it in the CMIS repository using the modeling tool. 
The modeling tool may expose the CMIS versioning capability.
Taking advantage of these capabilities, the business analysts may maintain multiple versions of the case model and may decide to deploy to a production system one of those versions.

In a system with no separation between design and runtime, a case worker may create a CMMN model starting from scratch or using a template stored in the CMIS repository.
In both cases, the resulting model may be stored in the CMIS repository for future usage as a template.
In systems with no clear separation between design and runtime, models will normally start incomplete and will evolve as the case workers process the instance.
These case models will continually evolve, and so, the version capabilities of CMIS will be used to keep track of the evolution of the model.

Eventually a case instance will be created and case workers will collaborate to complete the case using the client tool.
Documents of multiple types maybe required to process the case instance. 
For example, emails, word processing documents, spreadsheets, pictures, videos, voice recordings, case comments, etc.
Those documents will be stored in the case folder.
To organize those documents, the case workers may decide to create a folder structure under the case folder.
For example, it may be useful to create a sub-folder for correspondence.
That correspondence sub-folder may be further subdivided into an incoming correspondence sub-folder and an outgoing correspondence sub-folder.

In addition to the client tool that allows the case workers to interact with the case instance, other CMIS client programs could also interact with the case folder.
Documents in the case instance may be created by the case workers or it may be placed in the case instance by computer programs using the CMIS API to access the case file.
Events are raised when documents are added to the case, are modified, or are removed.
Because both documents and folders are \texttt{CaseFileItem}s, those events can be used in entry or exit criterion to tasks, stages, or milestones. 
So, as the case file is modified by either the case workers using the client tool or CMIS clients interact with the case file, then entry or exit criterion may be triggered.

\section{Conclusion}
\label{sec:Main-Conclusions}

This paper described how to implement the CMMN information model using  CMIS.
There is no need to extend CMIS to be used by CMMN, and only minor extensions to CMMN are proposed in this paper.
Two implementation alternatives were described. 
An \textit{integration} alternative where an external CMIS repository is used and an \textit{embedded} alternative where a CMIS repository is embedded within the CMMN engine. 
The \textit{integration} alternative will be appealing to process technology vendors, and the \textit{embedded} alternative will be appealing to content management vendors.
In both cases, the CMIS repository can be used to store the CMMN models to take advantage of CMIS versioning and meta-data.
Extensive sample Java pseudocode is provided and analysis of the meta-models was done to guide implementors.

\FloatBarrier
\bibliographystyle{abbrv}

\begin{thebibliography}{10}

\bibitem{Chemistry2014}
{Apache Chemistry}.
\newblock {Apache Chemistry}.
\newblock http://chemistry.apache.org/, 2015.
\newblock {Online; accessed 1-February-2015}.

\bibitem{OpenCMIS2015}
{Apache Chemistry}.
\newblock {Apache Chemistry OpenCMIS}.
\newblock http://chemistry.apache.org/java/opencmis.html, 2015.
\newblock {Online; accessed 1-February-2015}.

\bibitem{Brown2014}
J.~Brown and F.~Muller.
\newblock {\em {OpenCMIS Server Development Guide: Building custom CMIS servers
  with the Apache Chemistry OpenCMIS Server Framework}}.
\newblock https://github.com/cmisdocs/ServerDevelopmentGuidev2, 2nd edition
  edition, 2014.
\newblock {Online; accessed 18-December-2014}.

\bibitem{Clair2009}
L.~C. Clair, C.~Moore, and R.~Vitti.
\newblock {Dynamic Case Management - An Old Idea Catches New Fire}.
\newblock Technical report, Forrester, Cambridge, MA, 2009.

\bibitem{Ciccio2015}
C.~Di~Ciccio, A.~Marrella, and A.~Russo.
\newblock {Knowledge-Intensive Processes: Characteristics, Requirements and
  Analysis of Contemporary Approaches}.
\newblock {\em Journal on Data Semantics}, 4(1):29--57, 2015.

\bibitem{Hill2012}
J.~B. Hill.
\newblock {The Case for Case Management Solutions}.
\newblock Technical Report June, Gartner, 2012.

\bibitem{Marin2013cmmn}
M.~A. Marin, R.~Hull, and R.~Vacul\'{\i}n.
\newblock {Data Centric BPM and the Emerging Case Management Standard: A Short
  Survey}.
\newblock In M.~Rosa and P.~Soffer, editors, {\em Business Process Management
  Workshops}, volume 132, pages 24--30. Springer Berlin Heidelberg, Tallinn,
  Estonia, Sept. 2013.

\bibitem{Muller2013}
F.~Muller, J.~Brown, and J.~Potts.
\newblock {\em {CMIS and Apache Chemistry in Action}}.
\newblock Manning Publications Co., 2013.
\newblock ISBN 978-1-617-29115-9.

\bibitem{OASIS2012cmis}
OASIS.
\newblock {Content Management Interoperability Services (CMIS) Version 1.1}.
\newblock
  http://docs.oasis-open.org/cmis/CMIS/v1.1/csprd01/CMIS-v1.1-csprd01.pdf,
  2012.
\newblock {Online; accessed 1-February-2015}.

\bibitem{Omg2014cmmn}
OMG.
\newblock {Case Management Model and Notation, version 1.0}, 2014.
\newblock Document formal/2014-05-05.

\bibitem{Swenson2010}
K.~D. Swenson.
\newblock {\em {Mastering the Unpredictable}}.
\newblock Landmark Books. Meghan-Kiffer Press, Tampa, Florida, USA, 2010.

\bibitem{W3C2004Schema}
W3C.
\newblock {XML Schema Part 2: Datatypes}, October 2004.

\end{thebibliography}

\appendix
\section{CMMN and CMIS Meta-models}
\label{sec:App-metamodels}

The CMMN and the CMIS meta-models are provided here for reference purposes.
The four figures shown here have been copied from the formal specifications
\cite{Omg2014cmmn,OASIS2012cmis}.

\begin{figure}[b]
\centering
\includegraphics[keepaspectratio,width=6.5in]{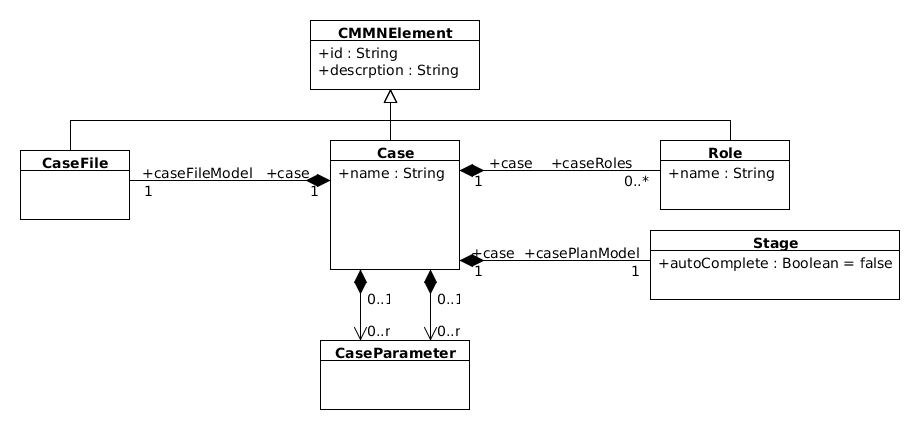}
\caption{CMMN High level meta-model}
\label{fig:CMMNhighlevel}
\end{figure}

\begin{figure}
\centering
\includegraphics[keepaspectratio,width=6.5in]{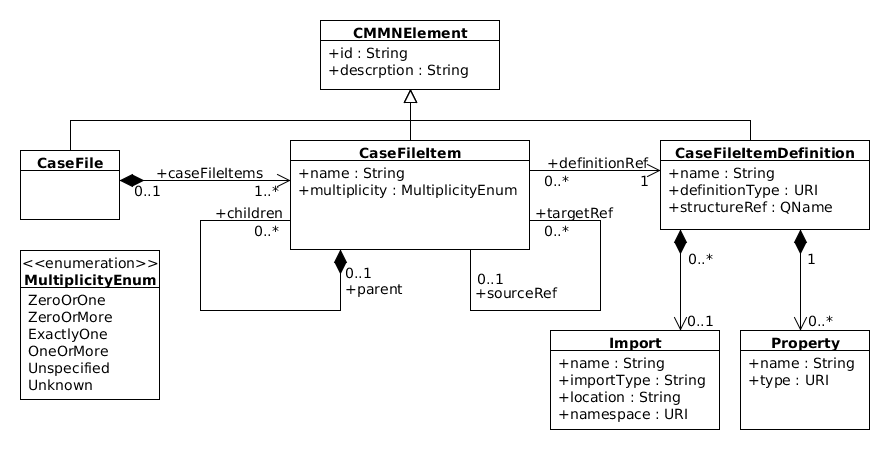}
\caption{CMMN case file item meta-model}
\label{fig:CMMNcasefile}
\end{figure}

Figure~\ref{fig:CMMNhighlevel} describes the CMMN high level meta-model showing the relationship between the \texttt{Case} and the \texttt{CaseFile} that implement the CMMN information model. 
Figure~\ref{fig:CMMNcasefile} describes how the \texttt{CaseFile} contains all the \texttt{CaseFileItem}s in the case.
In addition, it shows that \texttt{CaseFileItem}s can be used to create a folder structure using the composition relationship between \texttt{parent} and \texttt{children}; 
and it also shows that relationships between \texttt{CaseFileItem}s can be implemented using the reflexive association between \texttt{sourceRef} and \texttt{targetRef}.
These CMMN meta-models describe a CMMN model at modeling time and can be used for process interchange.

\begin{figure}
\centering
\includegraphics[keepaspectratio,width=6.5in]{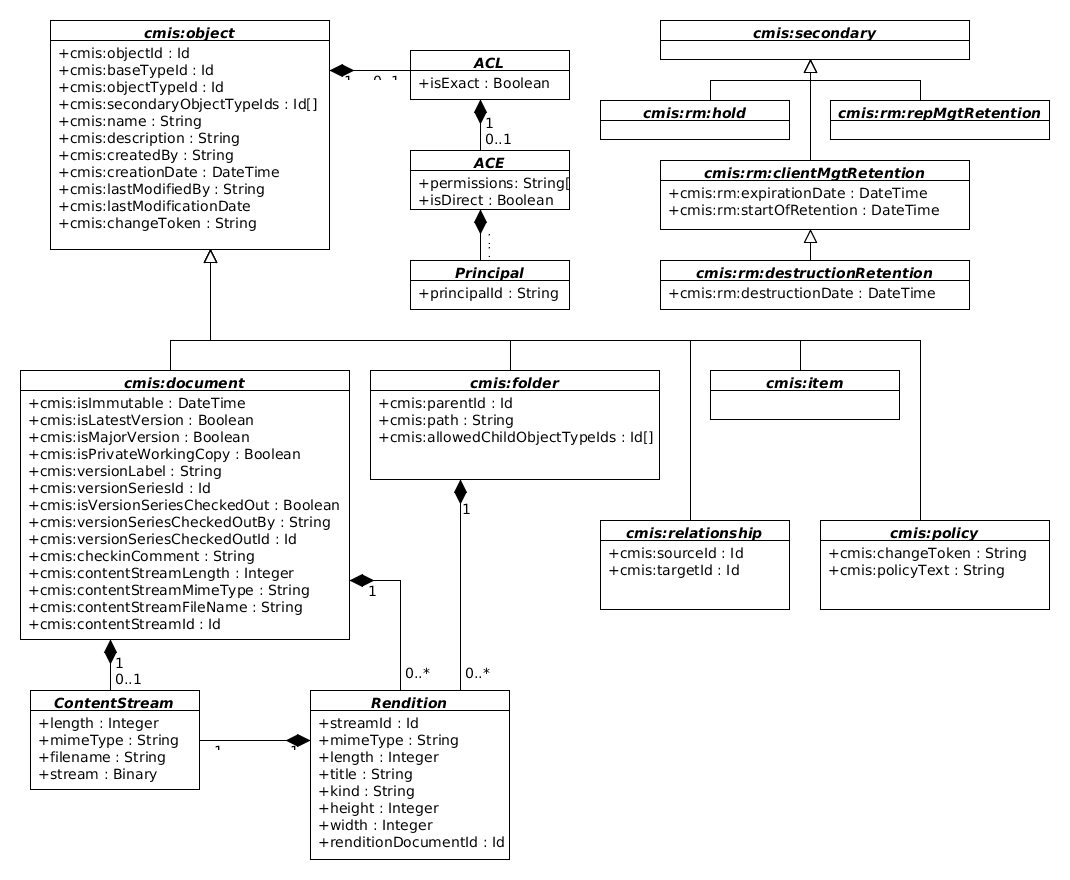}
\caption{CMIS meta-model}
\label{fig:CMISmetamodel}
\end{figure}

\begin{figure}
\centering
\includegraphics[keepaspectratio,width=6.5in]{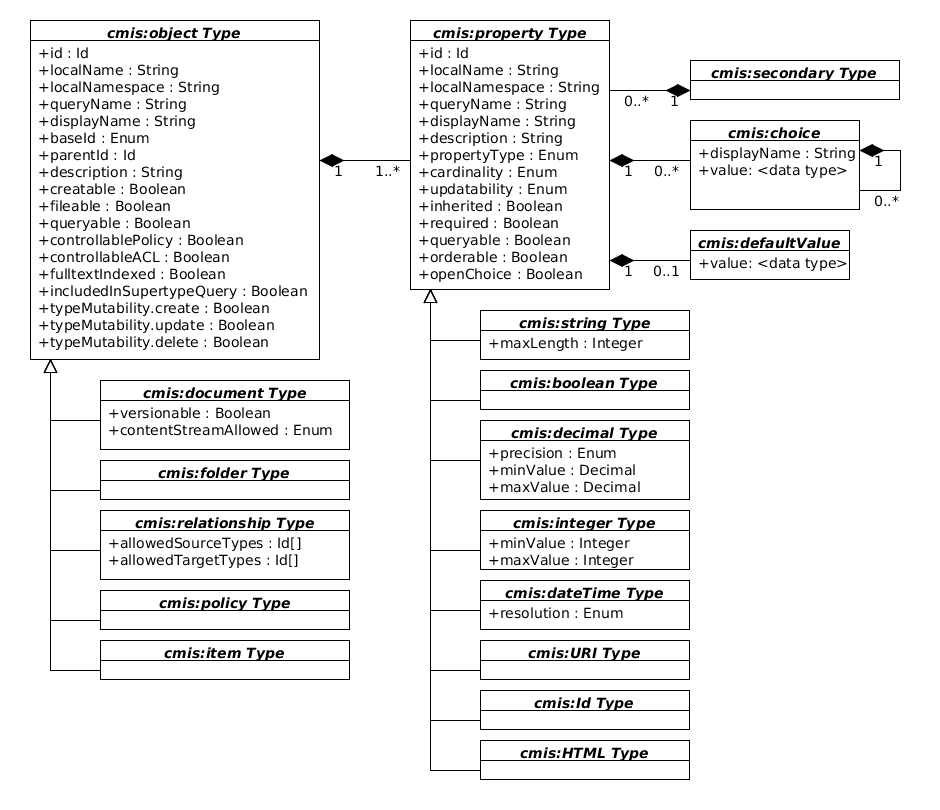}
\caption{CMIS Types}
\label{fig:CMIStypes}
\end{figure}

Figure~\ref{fig:CMISmetamodel} describes the CMIS objects meta-model, and 
Figure~\ref{fig:CMIStypes} describes the CMIS type system. 
These CMIS meta-models describe a content repository runtime, by describing the objects stored in the content repository at execution time.

\FloatBarrier

\lstset{
  language=Java,
  basicstyle=\footnotesize,
  breaklines=true,
  keepspaces=true,
  escapeinside={(*@}{@*)},
  numbers=left,
  numberstyle=\tiny\color{lgray},
  emph={CaseFileItemOperations,GetContentChangesForEventPropagation,PushChangeEvents,createCaseFileItemDocumentInstance,createCaseFileItemFolderInstance,createCaseFileItemRelationship,getCaseFileItemDocumentInstance,getCaseFileItemDocumentInstance2,getCaseFileItemDocumentInstanceChild,getCaseFileItemFolderInstance,getCaseFileItemFolderInstance2,getCaseFileItemFolderInstanceChild,getCaseFileItemInstance,getCaseFileItemInstance2,getCaseFileItemInstanceChild,getCaseFileItemInstanceParent,getCaseFileItemInstanceParents,getCaseFileItemInstanceSource,getCaseFileItemInstanceSourceOrTarget,getCaseFileItemInstanceTarget,getCaseItemInstanceProperty,getCaseItemInstancePropertySingleValue
},
  emphstyle=\underbar
  }

\section{Java pseudocode}
\label{sec:App-pseudocode}

All the sample Java pseudocode present here is uses Apache Chemistry OpenCMIS which is a standard CMIS reference client library for Java \cite{OpenCMIS2015}.
This pseudocode is an example of how to use OpenCMIS to implement the CMMN information model. 
It is not intended for production usage and so it lacks error recovery pseudocode.
There are few methods that use \texttt{System.out.println} in areas that are left as exercise to the reader to complete the methods.

This appendix lists the complete Java pseudocode in file \textsf{CaseFileItemOperations.java}. 
It starts as follows,

\begin{lstlisting}[firstline=2,lastline=40]
/*
* Copyright 2015 Jay Brown & Mike Marin
*
* Licensed under the Apache License, Version 2.0 (the "License");
* you may not use this file except in compliance with the License.
* You may obtain a copy of the License at
*
* http://www.apache.org/licenses/LICENSE-2.0
* Unless required by applicable law or agreed to in writing, software
* distributed under the License is distributed on an "AS IS" BASIS,
* WITHOUT WARRANTIES OR CONDITIONS OF ANY KIND, either express or implied.
* See the License for the specific language governing permissions and
* limitations under the License.
*
* It is part of a set of examples and is not intended for production use.
*
*/
package discovery.common;

import java.util.HashMap;
import java.util.List;
import java.util.Map;

import org.apache.chemistry.opencmis.client.api.ChangeEvent;
import org.apache.chemistry.opencmis.client.api.ChangeEvents;
import org.apache.chemistry.opencmis.client.api.CmisObject;
import org.apache.chemistry.opencmis.client.api.Document;
import org.apache.chemistry.opencmis.client.api.FileableCmisObject;
import org.apache.chemistry.opencmis.client.api.Folder;
import org.apache.chemistry.opencmis.client.api.ItemIterable;
import org.apache.chemistry.opencmis.client.api.ObjectId;
import org.apache.chemistry.opencmis.client.api.OperationContext;
import org.apache.chemistry.opencmis.client.api.Property;
import org.apache.chemistry.opencmis.client.api.Relationship;
import org.apache.chemistry.opencmis.client.api.Session;
import org.apache.chemistry.opencmis.commons.PropertyIds;
import org.apache.chemistry.opencmis.commons.data.ContentStream;
import org.apache.chemistry.opencmis.commons.enums.BaseTypeId;
import org.apache.chemistry.opencmis.commons.enums.VersioningState;
\end{lstlisting}

\subsection{CaseFile navigation operations}
\label{sec:ops}

This section describes the CMMN standard set of \texttt{CaseFileItem} operations 
for the behavioral model to navigate the information model
(see the CMMN specification \cite{Omg2014cmmn} section 7.3.1 CaseFileItem operations). 

The class \texttt{CaseFileItemOperations} is used to define all the methods described in this paper. 
The class constructor requires a CMIS session and a root folder that serves as the \texttt{CaseFile} for the case instance. 
Most of the methods operate on a case instance (\texttt{CaseFile}). 
For illustration purposes, some methods in this class can operate outside the case instance.

\begin{lstlisting}[firstnumber=last,firstline=42,lastline=73]
/**
 * CaseFileItemOperations source code example for CMIS.
 * 
 * Assumes that on CMIS server the class type for
 * org.apache.chemistry.opencmis.commons.enums.BaseTypeId has had the cmnn:index
 * property added. If you are using a subclass of document to represent our
 * CaseFiles then substitute your class type name for BaseTypeId.CMIS_DOCUMENT
 * in the code that follows.
 * 
 * 
 * All of this code is based on Apache Chemistry OpenCMIS objects. For the
 * JavaDocs for these objects please see
 * http://chemistry.apache.org/java/0.12.0/maven/apidocs/
 * 
 * 
 */
public class CaseFileItemOperations {

    private Folder rootCaseFolder; // initialized to a folder object
    private Session cmisSession; // initialized to a live session
    public static final String indexPropertyName = "cmnn:index";

    /**
     * @param rootCaseFolder
     *            - the root folder holding all artifacts for a case
     * @param cmisSession
     *            - live session to a CMIS server
     */
    public CaseFileItemOperations(Folder rootCaseFolder, Session cmisSession) {
        this.rootCaseFolder = rootCaseFolder;
        this.cmisSession = cmisSession;
    }
\end{lstlisting}

\subsubsection{\texttt{CaseFileItem} instances} \ 

The CMMN specification describes two overloaded operations to navigate \texttt{CaseFileItem} instances.
The first has a single input parameter (\texttt{itemName}).

\begin{alltt}
   getCaseFileItemInstance(\textbf{IN} itemName : String, 
                           \textbf{OUT} CaseFileItem instance)
\end{alltt}

Get a \texttt{CaseFileItem} (a \texttt{cmis:object} most likely a document or folder) instance with \texttt{itemName} (\texttt{cmis:name}) within the \texttt{CaseFile} container. 
If no \texttt{CaseFileItem} instance for the given \texttt{itemName} exists, 
an empty \texttt{cmis:document} (\texttt{CaseFileItem}) instance is returned.
If more than one \texttt{CaseFileItem} instance name has the same \texttt{itemName} (\texttt{cmis:name}), an arbitrary one should be returned.

This Java pseudocode provides three implementations for this operation. One returning a \texttt{cmis:object} (\texttt{getCaseFileItemInstance}), one returning a \texttt{cmis:document} (\texttt{getCaseFileItemDocumentInstance}), and finally one returning a \texttt{cmis:folder} (\texttt{getCaseFileItemFolderInstance}).

\begin{lstlisting}[firstnumber=last,firstline=75,lastline=112]
    /**
     * General version that returns any type of object Searches for documents
     * first. (folders, documents, items)
     * 
     * This is implemented using Query without a join so it will be compatible
     * with repositories that do not support joins. This could also be
     * implemented using folder.getDescendants(<depth>)
     * 
     * @param itemName
     * @return
     */
    public CmisObject getCaseFileItemInstance(String itemName) {
        return getCaseFileItemInstanceChild(this.rootCaseFolder, itemName);
    }

    /**
     * Find the first descendant document that matches the name supplied or null
     * if none found.
     * 
     * @param itemName
     * @return
     * 
     */
    public Document getCaseFileItemDocumentInstance(String itemName) {
        return getCaseFileItemDocumentInstanceChild(this.rootCaseFolder, itemName);
    }

    /**
     * Find the first descendant folder that matches the name supplied or null
     * if none found.
     * 
     * @param itemName
     * @return
     * 
     */
    public Folder getCaseFileItemFolderInstance(String itemName) {
        return getCaseFileItemFolderInstanceChild(this.rootCaseFolder, itemName);
    }
\end{lstlisting}

The second has two input parameters (\texttt{itemName} and \texttt{index}).

\begin{alltt}
   getCaseFileItemInstance(\textbf{IN}  itemName : String, 
                               index : Integer, 
                           \textbf{OUT} CaseFileItem instance)
\end{alltt}

Get a \texttt{CaseFileItem} (a \texttt{cmis:object} most likely a document or folder) instance with \texttt{itemName} (\texttt{cmis:name}) and \texttt{CaseFileItem}'s \texttt{index} (see Figure~\ref{fig:CaseFileIntegrated} and Figure~\ref{fig:CaseFileEmbeded}) within the \texttt{CaseFile} container. 
This operation is to be used for \texttt{CaseFileItem} (a \texttt{cmis:object} instances with a multiplicity greater than one. 
The \texttt{index} is used to identify a concrete \texttt{CaseFileItem} (a \texttt{cmis:object} most likely a document or folder)
instance from the collection of \texttt{CaseFileItem} instances. If no
\texttt{CaseFileItem} instance for the given \texttt{itemName} exists, or if the \texttt{index} is
out of the range of \texttt{CaseFileItem} instances, an empty \texttt{CaseFileItem} instance is returned.

Note that Java does not provide methods overloading, so a number 2 was appended to the method names. 
This Java pseudocode provides three implementations for this operation. One returning a \texttt{cmis:object} (\texttt{getCaseFileItemInstance2}), one returning a \texttt{cmis:document} (\texttt{getCaseFileItemDocumentInstance2}), and finally one returning a \texttt{cmis:folder} (\texttt{getCaseFileItemFolderInstance2}).

\begin{lstlisting}[firstnumber=last,firstline=114,lastline=230]
    /**
     * Return the first CmisObject that matches the itemName and the cmmn:index
     * or null if none found.
     * 
     * This is implemented using Query without a join so it will be compatible
     * with repositories that do not support joins. This could also be
     * implemented using folder.getDescendants(<depth>)
     * 
     * @param itemName
     * @param index
     * @return
     * 
     */
    public CmisObject getCaseFileItemInstance2(String itemName, Integer index) {

        CmisObject caseFileItem = null;
        OperationContext context = cmisSession.createOperationContext();
        context.setMaxItemsPerPage(100);
        context.setIncludeAllowableActions(Boolean.FALSE);

        // Note for this query to work you must have defined [indexPropertyName]
        // on your document and folder classes.
        // If you have your index field defined on a subclass of document/folder
        // change
        // the queries to select for your custom types
        String whereClause = "IN_TREE('" + this.rootCaseFolder.getId() 
                + "') AND cmis:name = '" + itemName + "' AND "
                + indexPropertyName + " = '" + index.toString() + "'";
        ItemIterable<CmisObject> queryResult = this.cmisSession.queryObjects(
                BaseTypeId.CMIS_DOCUMENT.value(), whereClause, true, context);

        // find the documents first
        if (queryResult.getPageNumItems() == 0) {
            for (CmisObject tempObj : queryResult) {
                caseFileItem = tempObj;
                break;
            }
        } else {
            // repeat for folder
            queryResult = this.cmisSession.queryObjects(BaseTypeId.CMIS_FOLDER.value(),
                whereClause, true, context);
            for (CmisObject tempObj : queryResult) {
                caseFileItem = tempObj;
                break;
            }
        }

        return caseFileItem;
    }

    /**
     * Return the first document that matches the itemName and the cmmn:index or
     * null if none found.
     * 
     * @param itemName
     * @param index
     * @return
     * 
     */
    public Document getCaseFileItemDocumentInstance2(String itemName, Integer index) {

        Document caseFileItem = null;
        OperationContext context = cmisSession.createOperationContext();
        context.setMaxItemsPerPage(100);
        context.setIncludeAllowableActions(Boolean.FALSE);

        // Note for this query to work you must have defined [indexPropertyName]
        // on your document class.
        // If you have your index field defined on a subclass of document change
        // this query to select for that type
        // instead of BaseTypeId.CMIS_DOCUMENT
        String whereClause = "IN_TREE('" + this.rootCaseFolder.getId() 
                + "') AND cmis:name = '" + itemName + "' AND "
                + indexPropertyName + " = '" + index.toString() + "'";
        ItemIterable<CmisObject> queryResult = this.cmisSession.queryObjects(
                BaseTypeId.CMIS_DOCUMENT.value(), whereClause, true, context);

        for (CmisObject tempObj : queryResult) {
            caseFileItem = (Document) tempObj;
            break;
        }
        return caseFileItem;
    }

    /**
     * Return the first folder that matches the itemName and the cmmn:index or
     * null if none found.
     * 
     * @param itemName
     * @param index
     * @return
     * 
     */
    public Folder getCaseFileItemFolderInstance2(String itemName, Integer index) {

        Folder caseFileItem = null;
        OperationContext context = cmisSession.createOperationContext();
        context.setMaxItemsPerPage(100);
        context.setIncludeAllowableActions(Boolean.FALSE);

        // Note for this query to work you must have defined [indexPropertyName]
        // on your folder class.
        // If you have your index field defined on a subclass of document change
        // this query to select for that type
        // instead of BaseTypeId.CMIS_DOCUMENT
        String whereClause = "IN_TREE('" + this.rootCaseFolder.getId() 
                + "') AND cmis:name = '" + itemName + "' AND "
                + indexPropertyName + " = '" + index.toString() + "'";
        ItemIterable<CmisObject> queryResult = this.cmisSession.queryObjects(
                BaseTypeId.CMIS_FOLDER.value(), whereClause, true, context);

        for (CmisObject tempObj : queryResult) {
            caseFileItem = (Folder) tempObj;
            break;
        }
        return caseFileItem;
    }
\end{lstlisting}

\subsubsection{\texttt{CaseFileItem} properties} \ 

\begin{alltt}
   getCaseFileItemInstanceProperty (\textbf{IN}  item : CaseFileItem instance,
                                        propertyName : String,
                                    \textbf{OUT} Element)
\end{alltt}

Get the value of a \texttt{CaseFileItem} instance property. 
If \texttt{propertyName} refers to a non-existing property of 
the \texttt{CaseFileItem} instance, an
empty \texttt{Element} MUST be returned. 
The Element returned MUST be of the specified property type for the \texttt{CaseFileItem} instance.

\begin{lstlisting}[firstnumber=last,firstline=232,lastline=257]
    /**
     * For retrieving a property from a fileInstance you only need to call the
     * .getProperty on the instance itself. As shown here in this method.
     * 
     * @param fileItemInstance
     * @param propertyName
     * @return OpenCMIS property object (see javadocs)
     */
    public <T> Property<T> getCaseItemInstanceProperty(Document fileItemInstance,
        String propertyName) {
        return fileItemInstance.getProperty(propertyName);
    }

    /**
     * Returns a single value property object by name or the first of a list of
     * properties
     * 
     * @param fileItemInstance
     * @param propertyName
     * @return Native value for property, i.e. String, Integer, Boolean, etc.,
     *         ...
     */
    public Object getCaseItemInstancePropertySingleValue(Document fileItemInstance,
        String propertyName) {
        return fileItemInstance.getProperty(propertyName).getFirstValue();
    }
\end{lstlisting}

\subsubsection{Using \texttt{CaseFileItem}s as folders}  \ 

The methods in this section are used to navigate \texttt{cmis:folders} when they implement the \texttt{CaseFileItem} self-referencing composition relationship between \texttt{parent} and \texttt{children} (see Figure~\ref{fig:CMMNcasefile}).

\begin{alltt}
   getCaseFileItemInstanceChild(\textbf{IN}  item : CaseFileItem instance,
                                    childName : String, 
                                \textbf{OUT} CaseFileItem)
\end{alltt}

Get a child \texttt{CaseFileItem} instance for a given \texttt{CaseFileItem} instance. 
This operation is valid for \texttt{CaseFileItem}s implemented as \texttt{cmis:folder}s (\texttt{cmis:folder}).  
The value of parameter \texttt{childName} specifies the name (\texttt{cmis:name}) of the child to get with in the \texttt{cmis:folder}. 
If no child of the given name exists for the \texttt{CaseFileItem} instance, an
empty \texttt{CaseFileItem} instance is returned.

This operation is provided to navigate the composition relationship between \texttt{CaseFileItem}s used to implement a folder structure.
They are represented in the CMMN meta-model (see \ref{fig:CMMNcasefile}) by the \texttt{parent} and \texttt{children} composition relationship. 
This operation navigates from the \texttt{parent} (always a \texttt{cmis:folder}) to the \texttt{child} (most likely a \texttt{cmis:document} or folder).

This Java pseudocode provides three implementations for this operation. One returning a \texttt{cmis:object} (\texttt{getCaseFileItemInstanceChild}), one returning a \texttt{cmis:document} (\texttt{getCaseFileItemDocumentInstanceChild}), and finally one returning a \texttt{cmis:folder} (\texttt{getCaseFileItemFolderInstanceChild}).

\begin{lstlisting}[firstnumber=last,firstline=259,lastline=366]
    /**
     * For the folder instance passed in, return any child object matching the
     * name
     * 
     * @param folderItemInstance
     * @param childName
     * @return 
     */
    public CmisObject getCaseFileItemInstanceChild(Folder folderItemInstance, 
        String childName) {
        CmisObject caseFileItem = null;
        OperationContext context = cmisSession.createOperationContext();
        context.setMaxItemsPerPage(100);
        context.setIncludeAllowableActions(Boolean.FALSE);

        // first search for documents. If you want to give preference to folders
        // move that code to run first.
        String whereClause = "IN_TREE('" + folderItemInstance.getId() 
                + "') AND cmis:name = '" + childName + "'";
        ItemIterable<CmisObject> queryResult = this.cmisSession.queryObjects(
                BaseTypeId.CMIS_DOCUMENT.value(), whereClause, true, context);

        if (queryResult.getPageNumItems() == 0) {
            for (CmisObject tempObj : queryResult) {
                System.out.println("Object name: " + tempObj.getName());
                caseFileItem = tempObj;
                break;
            }

        } else {
            // no documents found so let's do the same query for folders
            queryResult = this.cmisSession.queryObjects(BaseTypeId.CMIS_FOLDER.value(),
                whereClause, true, context);

            for (CmisObject tempObj : queryResult) {
                System.out.println("Object name: " + tempObj.getName());
                caseFileItem = tempObj;
                break;
            }
        }
        return caseFileItem;
    }

    /**
     * For the folder instance passed in, return any child folder matching the
     * name (any depth)
     * 
     * @param folderItemInstance
     * @param childName
     * @return
     * 
     */
    public Folder getCaseFileItemFolderInstanceChild(Folder folderItemInstance, 
        String childName) {
        Folder caseFileItem = null;
        OperationContext context = cmisSession.createOperationContext();
        context.setMaxItemsPerPage(100);
        context.setIncludeAllowableActions(Boolean.FALSE);

        String whereClause = "IN_TREE('" + folderItemInstance.getId() 
                + "') AND cmis:name = '" + childName + "'";
        ItemIterable<CmisObject> queryResult = this.cmisSession.queryObjects(
                BaseTypeId.CMIS_FOLDER.value(), whereClause, true, context);

        if (queryResult.getPageNumItems() == 0) {
            for (CmisObject tempObj : queryResult) {
                System.out.println("Object name: " + tempObj.getName());
                // Search was only for folders
                // so always ok to cast here
                caseFileItem = (Folder) tempObj;
                break;
            }
        }
        return caseFileItem;
    }

    /**
     * For the folder instance passed in, return any child document matching the
     * name (any depth)
     * 
     * @param folderItemInstance
     * @param childName
     * @return
     * 
     */
    public Document getCaseFileItemDocumentInstanceChild(Folder folderItemInstance, 
        String childName) {
        Document caseFileItem = null;
        OperationContext context = cmisSession.createOperationContext();
        context.setMaxItemsPerPage(100);
        context.setIncludeAllowableActions(Boolean.FALSE);

        String whereClause = "IN_TREE('" + folderItemInstance.getId() 
                + "') AND cmis:name = '" + childName + "'";
        ItemIterable<CmisObject> queryResult = this.cmisSession.queryObjects(
                BaseTypeId.CMIS_DOCUMENT.value(), whereClause, true, context);

        if (queryResult.getPageNumItems() == 0) {
            for (CmisObject tempObj : queryResult) {
                System.out.println("Object name: " + tempObj.getName());
                // Search was only for documents
                // so always ok to cast here
                caseFileItem = (Document) tempObj;
                break;
            }
        }
        return caseFileItem;
    }
\end{lstlisting}

\begin{alltt}
   getCaseFileItemInstanceParent(\textbf{IN}  item : CaseFileItem instance, 
                                 \textbf{OUT} CaseFileItem instance)
\end{alltt}

Get the parent \texttt{CaseFileItem} (\texttt{cmis:folder}) instance of a \texttt{CaseFileItem} instance.  
Note in the worse case, the parent will be the \texttt{CaseFile},
which is the parent of all the \texttt{CaseFileItem}s in a case.

This operation is provided to navigate the composition relationship between \texttt{CaseFileItem}s used to implement a folder structure.
They are represented in the CMMN meta-model (see \ref{fig:CMMNcasefile}) by the \texttt{parent} and \texttt{children} composition relationship. 
This operation navigates from the \texttt{child} (most likely a \texttt{cmis:document} or folder) to the \texttt{parent} (always a \texttt{cmis:folder}).

\begin{lstlisting}[firstnumber=last,firstline=368,lastline=390]
    /**
     * Takes any CMIS object (document or folder) and returns its first parent.
     * If this is a folder it can only have one parent. If it is a document it
     * may have multiple parents. (if multifiled)
     * 
     * @param caseFileItem
     * @return
     */
    public Folder getCaseFileItemInstanceParent(FileableCmisObject caseFileItem) {
        return caseFileItem.getParents().get(0);
    }

    /**
     * Takes any CMIS object (document or folder) and returns list of parents.
     * If this is a folder it can only have one parent. If it is a document it
     * may have multiple parents. (if multifiled)
     * 
     * @param caseFileItem
     * @return 
     */
    public List<Folder> getCaseFileItemInstanceParents(FileableCmisObject caseFileItem) {
        return caseFileItem.getParents();
    }
\end{lstlisting}

\subsubsection{Relationships between \texttt{CaseFileItem}s}   \ 

The methods in this section are used to navigate the  \texttt{cmis:relationship} used to implement the \texttt{CaseFileItem} self-referencing reflexive association between \texttt{sourceRef} and \texttt{targetRef} (see Figure~\ref{fig:CMMNcasefile}).

\begin{alltt}
   getCaseFileItemInstanceSource(\textbf{IN}  item : CaseFileItem instance, 
                                 \textbf{OUT} CaseFileItem instance)
\end{alltt}

Get the source \texttt{CaseFileItem} instance of a \texttt{CaseFileItem} instance. 

This operation is provided to navigate relationships between \texttt{CaseFileItem}s.
They are represented in the CMMN meta-model (see \ref{fig:CMMNcasefile}) by the \texttt{sourceRef} and \texttt{TargetRef} relationship. 
This operation navigates from the \texttt{targetRef} to the \texttt{sourceRef}.

\begin{lstlisting}[firstnumber=last,firstline=392,lastline=418]
    /**
     * Return all cmis:relationship items associated with the supplied
     * caseFileItem where the *source* is the object specified
     * 
     * @param caseFileItem
     */
    public ItemIterable<CmisObject> getCaseFileItemInstanceSource(
        FileableCmisObject caseFileItem) {

        OperationContext context = cmisSession.createOperationContext();
        context.setMaxItemsPerPage(100);
        context.setIncludeAllowableActions(Boolean.FALSE);

        String whereClause = "cmis:sourceId = '" + caseFileItem.getId() + "'";

        // Query all relationships (even if relationship is for item outside of
        // this case)
        // this query requires that the repository have
        // 1) cmis:relationship objects
        // 2) cmis:relationship object with queryable=true
        ItemIterable<CmisObject> queryResult = this.cmisSession.queryObjects(
                BaseTypeId.CMIS_RELATIONSHIP.value(), whereClause, true, context);

        // any additional filtering or processing you want to do on the list of
        // relationships here...
        return queryResult;
    }
\end{lstlisting}

\begin{alltt}
   getCaseFileItemInstanceTarget(\textbf{IN}  item : CaseFileItem instance,
                                     targetName : String, 
                                 \textbf{OUT} CaseFileItem instance)
\end{alltt}

Get a target \texttt{CaseFileItem} instance for a given \texttt{CaseFileItem} instance. 
The value of parameter \texttt{childName} specifies the name (\texttt{cmis:name}) of the target to get. 
If no target of the given name exists for the \texttt{CaseFileItem} instance, an
empty \texttt{CaseFileItem} instance will be returned.

This operation is provided to navigate relationships between \texttt{CaseFileItem}s.
They are represented in the CMMN meta-model (see \ref{fig:CMMNcasefile}) by the \texttt{sourceRef} and \texttt{targetRef} relationship. 
This operation navigates from the \texttt{sourceRef} to the \texttt{targetRef}.

\begin{lstlisting}[firstnumber=last,firstline=420,lastline=475]
    /**
     * Return all cmis:relationship items associated with the supplied
     * caseFileItem where the *target* is the object specified
     * 
     * @param caseFileItem
     */
    public ItemIterable<CmisObject> getCaseFileItemInstanceTarget(
        FileableCmisObject caseFileItem) {

        OperationContext context = cmisSession.createOperationContext();
        context.setMaxItemsPerPage(100);
        context.setIncludeAllowableActions(Boolean.FALSE);

        String whereClause = "cmis:targetId = '" + caseFileItem.getId() + "'";

        // Query all relationships (even if relationship is for item outside of
        // this case)
        // this query requires that the repository have
        // 1) cmis:relationship objects
        // 2) cmis:relationship object with queryable=true
        ItemIterable<CmisObject> queryResult = this.cmisSession.queryObjects(
                BaseTypeId.CMIS_RELATIONSHIP.value(), whereClause, true, context);

        // any additional filtering or processing you want to do on the list of
        // relationships here...
        return queryResult;
    }

    /**
     * Return all cmis:relationship items associated with the supplied
     * caseFileItem.
     * 
     * @param caseFileItem
     */
    public ItemIterable<CmisObject> getCaseFileItemInstanceSourceOrTarget(
        FileableCmisObject caseFileItem) {

        OperationContext context = cmisSession.createOperationContext();
        context.setMaxItemsPerPage(100);
        context.setIncludeAllowableActions(Boolean.FALSE);

        String whereClause = "cmis:sourceId = '" + caseFileItem.getId() 
                + "' OR (cmis:targetId = '" + caseFileItem.getId() + "')";

        // Query all relationships (even if relationship is for item outside of
        // this case)
        // this query requires that the repository have
        // 1) cmis:relationship objects
        // 2) cmis:relationship object with queryable=true
        ItemIterable<CmisObject> queryResult = this.cmisSession.queryObjects(
                BaseTypeId.CMIS_RELATIONSHIP.value(), whereClause, true, context);

        // any additional filtering or processing you want to do on the list of
        // relationships here...
        return queryResult;
    }
\end{lstlisting}

\subsection{CaseFile modification operations}
\label{sec:other}

This section shows some examples on how to use CMIS to modify the case instance (\texttt{CaseFile}) information model. 
Three creation methods are included here, two of them allow to create folders and documents in the root folder representing the case instance (\texttt{CaseFile}), and one to create relationships between CMIS objects. They can be used as examples of how the case information model can be modified.

Updates and deletions of objects in the case information model can be easily done using standard OpenCMIS \cite{OpenCMIS2015} method calls in the corresponding objects.

\begin{lstlisting}[firstnumber=last,firstline=477,lastline=622]
    /**
     * Create a new caseFileItemFolder instance
     * 
     * @param folderItemInstance
     *            - parent folder item for this new folder (optional) If omitted
     *            folder will be created as child of root caseItemInstance for
     *            this case.
     * @param itemFolderName
     *            - name of the new subfolder
     * @param objectType
     *            - (optional) specific subtype of cmis:folder to create If
     *            omitted cmis:folder will be used
     * @param additionalProperties
     *            (optional) any other optional properties to set on the object
     * 
     * @return new caseFileItemFolder instance
     */
    public Folder createCaseFileItemFolderInstance(Folder folderItemInstance, 
            String itemFolderName, String objectType,
            Map<String, Object> additionalProperties) {

        if (folderItemInstance == null) {
            folderItemInstance = this.rootCaseFolder;
        }

        if (additionalProperties == null) {
            // create a properties object
            additionalProperties = new HashMap<String, Object>();
        }

        // set the object type (specific subclass of cmis:relationship) - not
        // required
        if (objectType != null) {
            additionalProperties.put(PropertyIds.OBJECT_TYPE_ID, objectType);
        } else {
            additionalProperties.put(PropertyIds.OBJECT_TYPE_ID, 
                BaseTypeId.CMIS_FOLDER.value());
        }

        // set the name (required)
        additionalProperties.put(PropertyIds.NAME, itemFolderName);

        return folderItemInstance.createFolder(additionalProperties);
    }

    /**
     * Create a new caseFileItemDocument instance
     * 
     * @param folderItemInstance
     *            - parent folder item for this new document (optional) If
     *            omitted document will be created as child of root
     *            caseItemInstance for this case.
     * @param itemFolderName
     *            - name of the new document
     * @param objectType
     *            - (optional) specific subtype of cmis:document to create If
     *            omitted cmis:document will be used
     * @param additionalProperties
     *            (optional) any other optional properties to set on the object
     * @param versioningState
     *            (optional) initial versioning state of the document.
     * @param contentStream
     *            (optional) content Stream for the document. If omitted then
     *            document will contain only metadata (see additionalProperties)
     * @param additionalProperties
     * 
     * @return new caseItemDocument instance
     */
    public Document createCaseFileItemDocumentInstance(Folder folderItemInstance, 
            String itemDocumentName,
            String objectType, VersioningState versioningState, ContentStream contentStream,
            Map<String, Object> additionalProperties) {

        if (folderItemInstance == null) {
            folderItemInstance = this.rootCaseFolder;
        }

        if (additionalProperties == null) {
            // create a properties object
            additionalProperties = new HashMap<String, Object>();
        }

        // set the object type (specific subclass of cmis:relationship) - not
        // required
        if (objectType != null) {
            additionalProperties.put(PropertyIds.OBJECT_TYPE_ID, objectType);
        } else {
            additionalProperties.put(PropertyIds.OBJECT_TYPE_ID, 
                BaseTypeId.CMIS_DOCUMENT.value());
        }

        // set the name (required)
        additionalProperties.put(PropertyIds.NAME, itemDocumentName);

        return folderItemInstance.createDocument(additionalProperties, contentStream, 
                   versioningState);
    }

    /**
     * Create a new relationship object and return it.
     * 
     * @param source
     *            - source object to set on relationship
     * @param target
     *            - target object to set on relationship
     * @param objectType
     *            - optional - specific subclass of relationship required
     * @param additionalProperties
     *            - additional properties to set on the relationship - these
     *            must all be defined on the objectType you specify.
     * 
     * @return The CMIS ObjectId of the new relationship object.
     */
    public ObjectId createCaseFileItemRelationship(CmisObject source, CmisObject target, 
            String objectType,
            Map<String, String> additionalProperties) {
        if (additionalProperties == null) {
            // create a properties object
            additionalProperties = new HashMap<String, String>();
        }

        // set the object type (specific subclass of cmis:relationship) - not
        // required
        if (objectType != null) {
            additionalProperties.put(PropertyIds.OBJECT_TYPE_ID, objectType);
        } else {
            additionalProperties.put(PropertyIds.OBJECT_TYPE_ID, 
                BaseTypeId.CMIS_RELATIONSHIP.value());
        }

        // set the source and target (required)
        additionalProperties.put(PropertyIds.SOURCE_ID, source.getId());
        additionalProperties.put(PropertyIds.TARGET_ID, target.getId());

        return cmisSession.createRelationship(additionalProperties);
    }

    /**
     * To update a folder or document use the object's CmisObject
     * updateProperties(Map<String,?> properties) method directly.
     * 
     * To delete use the object's .delete() method.
     * 
     * See http://chemistry.apache.org/java/0.12.0/maven/apidocs/ for more
     * details on the signatures of these methods.
     */
\end{lstlisting}

\subsection{Event propagation}
\label{sec:event}

This section describes how to receive the events from the CMIS repository.
The following methods are included in this class for illustration purposes, but these methods are not case instance specific. 
They will receive events from all the case instances in the CMIS repository.
These methods should be executed in their own thread, because \texttt{GetContentChangesForEventPropagation} will go into a infinite loop.
Most implementations will encapsulate the two methods shown in this section in another class to be executed in its own thread.

\begin{lstlisting}[firstnumber=last,firstline=624]
    /**
     * *************************************************************************
     * ******** This following section is code to demonstrate event push
     * techniques for CMIS.
     * 
     * This code would be put into its own module where it could run
     * continuously
     * 
     * *************************************************************************
     * ********
     * 
     */

    /**
     * GetContentChangesForEventPropagation Shows how to poll/push changes to
     * clients who expect push notifications of document change events.
     * 
     * @param session
     *            - active OpenCMIS session with CMIS server
     * @throws InterruptedException
     *             due to use of Thread.sleep
     */
    public void GetContentChangesForEventPropagation(Session session) 
        throws InterruptedException {

        Integer BatchTimeIntervalInMS = 1000; // create a batch of changes every
                                              // second
        long maxChangesToProcessInSingleBatch = 500; // if you need to limit
                                                     // batch size

        // if your repository supports property details with content changes
        // then you may set this to true (may affect performance)
        Boolean includePropertiesWithChangeEvents = false;

        // determine starting point (token) for changes
        String latestChangeToken = session.getRepositoryInfo().getLatestChangeLogToken();
        String previousChangeToken = "";
        System.out.println("Using initial change token:" + latestChangeToken);

        // main event processing loop - infinite
        while (true) {
            Thread.sleep(BatchTimeIntervalInMS);

            ChangeEvents changeEvents = session.getContentChanges(latestChangeToken, 
                    includePropertiesWithChangeEvents, maxChangesToProcessInSingleBatch);

            // push these changes to subscribed clients if there are new events
            // to process
            if (changeEvents.getChangeEvents().size() > 0) {
                // The changelog token can be the same and there still might
                // be one event in the list. This is so that clients can
                // validate that there is overlap and thus that they have not
                // missed an event. ( verify there is no gap)
                if (!previousChangeToken.equals(latestChangeToken)) {
                    PushChangeEvents(changeEvents);
                } else {
                    System.out.println(
                        "Duplicate overlap that we don't have to process found.");
                }
            } else {
                System.out.println("No changes found.");
            }

            // update our change token based on the last change we processed
            previousChangeToken = latestChangeToken;
            latestChangeToken = changeEvents.getLatestChangeLogToken();
            System.out.println("updated changeToken:" + latestChangeToken);

        }
    }

    /**
     * 
     * @param changeEvents
     *            - openCMIS ChangeEvents holder
     *            (org.apache.chemistry.opencmis.client.api.ChangeEvents)
     */
    public void PushChangeEvents(ChangeEvents changeEvents) {
        // ... impl dependent code to push these events here

        for (ChangeEvent ce : changeEvents.getChangeEvents()) {

            // print statements just for debugging
            System.out.println("ID: " + ce.getObjectId());
            System.out.println("type: " + ce.getChangeType());

            // process each change event separately or in a batch
            // using whatever push technology your system requires. (e.g. Comet)
        }
    }
}   // End class CaseFileItemOperations
\end{lstlisting}

\end{document}